\documentclass[showpacs,aps,graphicx,twocolumn]{revtex4}
\usepackage{amssymb}
\usepackage{amsmath}
\usepackage{graphicx}
\usepackage{array}
\usepackage{multirow}

\begin{document}

\title{Efficient multipartite entanglement purification with the entanglement link from a subspace\footnote{Published in Phys. Rev. A \textbf{84}, 052312  (2011)}}

\author{Fu-Guo Deng\footnote{Corresponding author:
fgdeng@bnu.edu.cn} }
\address{  Department of Physics, Applied Optics Beijing Area Major Laboratory, Beijing Normal University, Beijing 100875, China }
\date{\today }

\begin{abstract}
We present an efficient multipartite entanglement purification
protocol (MEPP) for $N$-photon systems in a
Greenberger-Horne-Zeilinger state with parity-check detectors. It
contains two parts. One is the conventional MEPP with which the
parties can obtain a high-fidelity  $N$-photon ensemble directly,
similar to the MEPP with controlled-not gates. The other is our
recycling MEPP in which the entanglement link is used to produce
some $N$-photon entangled systems from entangled $N'$-photon
subsystems ($2 \leq N'<N$) coming from the instances which are just
discarded in all existing conventional MEPPs. The entangled
$N'$-photon subsystems are obtained efficiently by measuring the
photons with potential bit-flip errors. With these two parts, the
present MEPP has a higher efficiency than all other conventional
MEPPs.
\end{abstract}
 \maketitle

\section{introduction}

Entanglement is an important quantum resource for quantum
communication and computation \cite{book}. The powerful speedup of
quantum computation resorts to multipartite entanglement
\cite{book}. Some branches in quantum communication should require
entanglement to set up the quantum channel. For example, quantum
teleportation \cite{teleportation} requires a maximally entangled
photon pair to set up the quantum channel for the teleportation of
an unknown single-qubit state. Some quantum key distribution (QKD)
protocols work with maximally entangled photon systems
\cite{Ekert91,BBM92,LongLiu,lixhpra,QKDdevice}. Moreover, people
should resort to quantum repeaters for a long-distance QKD or a QKD
network as the quantum signals (that is,  single photons
\cite{bb84,QOTP}, weak pulses \cite{faint1,faint2,faint3,faint4}, or
entangled photons \cite{Ekert91,BBM92,LongLiu,lixhpra,QKDdevice})
can only be transmitted over a fiber or a free space not more than
several hundreds kilometers with current technology (for example, an
experimental demonstration of free-space decoy QKD over 144 km was
obtained in Ref. \cite{QKDexp}). In a quantum repeater, entanglement
is required to connect two neighboring nodes. The high capacity of
quantum dense coding \cite{densecoding,densecoding2,densecoding3}
should also resort to maximally entangled photon systems. Quantum
secret sharing (QSS) \cite{QSS,QSS2,QSS3} and quantum state sharing
(QSTS) \cite{QSTS,QSTS2,QSTS3,QSTS5} require the parties in quantum
communication possess maximally entangled multi-photon systems.
However, entangled photon systems inevitably suffer from channel
noise when the entangled photons propagate away from each other. For
instance, the thermal fluctuation, vibration, the imperfection of an
optical fiber, and the birefringence effects will inevitably affect
the polarization of photons. In general, the interaction will make
an entangled system be in a less entangled state or even in a mixed
state. The decoherence of entanglement in quantum systems will
affect quantum communication largely. It will decrease the security
of QKD, QSS, and QSTS protocols if a maximally entangled state
transmitted over a noisy channel becomes a mixed entangled state as
a vicious eavesdropper can exploit the decoherence to hide her
illegal action. The non-maximally entangled quantum channel will
decrease the fidelity of quantum teleportation and quantum dense
coding.

Entanglement purification
\cite{Bennett1996,Deutsch,Pan1,Simon,shengpra2008,shengpratwostep,shengpraonestep,lixhepp,dengonestep}
is a useful tool for the parties in quantum communication to  obtain
some maximally entangled photon pairs from a set of less-entangled
photon pairs with the help of local operations and classical
communications. In 1996, Bennett \emph{et al}. \cite{Bennett1996}
proposed the first entanglement purification protocol (EPP) to
purify a Werner state \cite{werner}, resorting to quantum
controlled-not (CNOT) gates. Subsequently, Deutsh \emph{et al}.
\cite{Deutsch} optimized this EPP with  two additional specific
unitary operations. In 2001, Pan \emph{et al}. \cite{Pan1}
introduced an EPP with linear optical elements and an ideal
entanglement source by sacrificing a half of the efficiency. In
2002, Simon and Pan \cite{Simon} proposed an EPP with a currently
available parametric down-conversion (PDC) source. In 2008, an
efficient EPP \cite{shengpra2008} based on a PDC source was proposed
with cross-Kerr nonlinearity. It has the same efficiency as the EPP
by Bennett \emph{et al.} with perfect CNOT gates. In 2010, the
concept of deterministic entanglement purification was proposed
\cite{shengpratwostep} for two-photon entangled systems, which is
far different from the conventional entanglement purification
protocols (CEPPs) \cite{Bennett1996,Deutsch,Pan1,Simon,shengpra2008}
as the former works in a deterministic way, while the latter works
in a probabilistic way. In 2010, we introduced a two-step
deterministic entanglement purification protocol (DEPP)
\cite{shengpratwostep} for entangled photon pairs, resorting to
hyerentanglement. Subsequently, a one-step DEPP
\cite{shengpraonestep,lixhepp} was proposed, only resorting to the
spatial entanglement of a practical PDC source and linear optical
elements. In essence, both the CEPPs
\cite{Bennett1996,Deutsch,Pan1,Simon,shengpra2008} and the DEPPs
\cite{shengpratwostep,shengpraonestep,lixhepp,dengonestep} are based
on entanglement transfer. The CEPPs are based on the entanglement
transfer between different entangled photon systems, while the DEPPs
are based on the transfer between different degrees of freedom of an
entangled photon system itself.

By far, there have been several interesting EPPs
\cite{Bennett1996,Deutsch,Pan1,Simon,shengpra2008,shengpratwostep,shengpraonestep,lixhepp}
focusing on   entangled two-photon systems, while there are only two
EPPs for multipartite photon systems \cite{Murao,shengepjd} and an
EPP for multipartite electronic systems \cite{shengpla} with charge
detection. In 1998, Murao \emph{et al.} \cite{Murao} proposed a
multipartite entanglement purification protocol (MEPP) to purify
multipartite quantum systems in a Greenberger-Horne-Zeilinger (GHZ)
with CNOT gates, following some ideas in the EPP by Bennett \emph{et
al.} \cite{Bennett1996}. In 2009, a MEPP based on cross-Kerr
nonlinearities was proposed \cite{shengepjd}. In this protocol, the
cross-Kerr nonlinearity is used to construct a nondestructive
quantum nondemolition detector (QND) \cite{QND1} which has the
functions of both a parity-check gate and  a photon-number detector.
With QNDs, the parties can obtain some high-fidelity GHZ states from
an ensemble in a mixed entangled state by performing this MEPP
repeatedly. In both these two MEPPs, the original fidelity before
the MEPPs is required to be larger than $1/2$ and a lot of entangled
quantum resource is discarded. So does the MEPP for electronic
systems \cite{shengpla}.

In this article, we will present an efficient MEPP for $N$-photon
systems in a GHZ state. It contains two parts. One is our
conventional MEPP with which the parties can obtain a high-fidelity
$N$-photon ensemble directly, similar to the conventional MEPP with
controlled-not gates \cite{Murao}, but it doubles the efficiency of
the MEPP with cross-Kerr nonlinearity in Ref. \cite{shengepjd} and
the MEPP for electronic systems \cite{shengpla}. The other is our
recycling MEPP in which the entanglement link is used to produce
some $N$-photon entangled systems from subspaces. That is, the
parties in quantum communication first distil some entangled
$N'$-photon subsystems ($2 \leq N'<N$) from the cross-combinations
which are just the discarded instances in the conventional MEPPs
\cite{Murao,shengepjd,shengpla}, and then they produce some
$N$-photon entangled systems with entanglement link. It is
interesting to show that the entangled $N'$-photon subsystems are
obtained efficiently by measuring the potential photons with
bit-flip errors in the two cross-combinations of two $N$-photon
states. We discuss the entanglement link in detail for the
three-photon systems. Moreover, the present MEPP works by replacing
parity-check detectors with CNOT gates.  With these two parts, the
present MEPP has a higher efficiency than all other conventional
MEPPs \cite{Murao,shengepjd,shengpla}.

This article is organized as follows: we  discuss our conventional
three-photon entanglement purification for bit-flip errors in Sec.
\ref{seciib}. In Sec. \ref{seciic}, we give the detail for the
two-photon entanglement purification from three-photon systems with
potential bit-flip errors. That is, how can the parties obtain a
high-fidelity entangled two-photon subsystem from the
cross-combinations of two three-photon systems which are just
discarded in other conventional MEPPs
\cite{Murao,shengepjd,shengpla}. In Sec. \ref{seciid}, we give a way
for three-photon entanglement production from two-photon subsystems
with entanglement link. The differences of the efficiency and the
fidelity between the present MEPP and the conventional MEPPs are
shown in Sec. \ref{seciie}. In Sec. \ref{seciii}, a conventional
three-photon entanglement purification for phase-flip errors is
given.  A discussion and a summary are given in Sec. \ref{secv}. In
Appendix \ref{seciv}, we exploit four-photon systems as an example
to describe the principle of the present MEPP for $N$-photon
systems.

\section{High-efficiency three-photon entanglement purification for bit-flip errors with entanglement link}

\label{secii}

\subsection{Parity-check detector based on cross-Kerr nonlinearity}

\label{seciia}

Cross-Kerr nonlinearity is a powerful tool for us to construct QNDs
\cite{QND1,QND2}.  The cross-kerr nonlinearity has been used to
prepare  CNOT gates \cite{QND1} and complete a local Bell-state
analysis \cite{shengpraCHBS,QND2}. Also it can be used to fulfill
the quantum entanglement purification protocols
\cite{shengpra2008,shengpratwostep,shengepjd,wangcqic} and the
entanglement concentration protocol \cite{shengpraEC}. The
Hamiltonian of the cross-Kerr nonlinearity is $H_{ck}=\hbar\chi
a^{+}_{s}a_{s}a^{+}_{p}a_{p}$ \cite{QND1,QND2}. Here $a^{+}_{s}$ and
$a^{+}_{p}$ are the creation operations, and $a_{s}$ and $a_{p}$ are
the destruction operations. $\chi$ is the coupling strength of the
nonlinearity, which is decided by the property of nonlinear
material. Suppose a signal state
$|\Psi\rangle_s=c_{0}|0\rangle_{s}+c_{1}|1\rangle_{s}$
($|0\rangle_{s}$ and $|1\rangle_{s}$ denote that there are no photon
and one photon respectively in this state) and a coherent probe beam
in the state $|\alpha\rangle$ couple with a cross-Kerr nonlinearity
medium, the whole system evolves as:
\begin{eqnarray}
U_{ck}|\Psi\rangle_{s}|\alpha\rangle_{p}&=&
e^{iH_{ck}t/\hbar}[c_{0}|0\rangle_{s}+c_{1}
|1\rangle_{s}]|\alpha\rangle_{p} \nonumber\\
&=& c_{0}|0\rangle_{s}|\alpha\rangle_{p}+c_{1}|1\rangle_{s}| \alpha
e^{i\theta}\rangle_{p},
\end{eqnarray}
where $\theta=\chi t$ and $t$ is the interaction time. The coherent
beam picks up a phase shift $\theta$ directly proportional to the
number of the photons in the Fock state $|\Psi\rangle_s$, which can
be read out with  a general homodyne-heterodyne measurement. So one
can exactly check the number of photons in the Fock state but not
destroy them. We will exploit this feature to construct our QND in
our EPP, instead of the CNOT gates in the CEPPs
\cite{Bennett1996,Deutsch,Murao}.

\begin{figure}[!h]
\begin{center}
\includegraphics[width=6cm,angle=0]{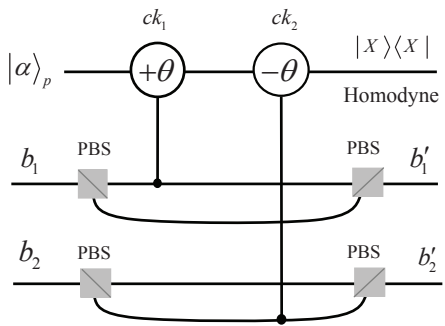}
\caption{The principle of the nondestructive quantum nondemolition
detector (QND) in the present MEPP. PBS represents a polarizing beam
splitter which transmits the $|H\rangle$ polarization photons and
reflect the $|V \rangle$ polarization photons. $\pm \theta$
represent two cross-Kerr nonlinear media which introduce the phase
shifts $\pm \theta$ when there is a photon passing through the
media.} \label{fig1_QND}
\end{center}
\end{figure}

The principle of our QND is shown in Fig.\ref{fig1_QND}, similar to
those in Refs. \cite{QND1,shengpraCHBS}. It is composed of two
cross-Kerr nonlinearities ($ck_1$ and $ck_2$), four polarization
beam splitters (PBSs), a coherent beam $\vert \alpha\rangle_p$, and
an X quadrature measurement. $b_1$ and $b_2$ represent the up
spatial mode and the down spatial mode, respectively.  Each
polarization beam splitter (PBS) is used to pass through the
horizontal polarization photons $|H\rangle$ and reflect the vertical
polarization photons $|V\rangle$. The cross-Kerr nonlinearity will
make the coherent beam $\vert \alpha \rangle_p$ pick up a phase
shift $\theta$ or $-\theta$ if there is a photon in the mode. The
probe beam $\vert \alpha \rangle_p$ will pick up a phase shift
$\theta$ or $-\theta$ if the state of the two photons injected into
the two spatial modes $b_1$ and $b_2$ is $|HH\rangle_{b_1b_2}$ or
$|VV\rangle_{b_1b_2}$, respectively; otherwise it picks up a phase
shift $0$ when the state of the two photons injected into the two
spatial modes $b_1$ and $b_2$ is $|VH\rangle_{b_1b_2}$ or
$|HV\rangle_{b_1b_2}$. That is, when the parity of the two photons
is even, the coherent beam $\vert \alpha \rangle_p$ will pick up a
phase shift $\theta$ or  $-\theta$; otherwise it will pick up  a
phase shift $0$. Each party of quantum communication can determine
the parity of his two photons with an X quadrature measurement in
which the the states $\vert \alpha e^{\pm i\theta} \rangle_p$ cannot
be distinguished \cite{QND1,QND3}. With this QND, each party can
distinguish superpositions and mixtures of $|HH\rangle$ and
$|VV\rangle$ from $|HV\rangle$ and $|VH\rangle$.

\subsection{Conventional three-photon  entanglement purification for bit-flip errors}
\label{seciib}

For three-photon systems, there are eight GHZ states for
polarization degree of freedom. They can be written as follows
\begin{eqnarray}
|\Phi_{0}^{\pm}\rangle_{ABC}=\frac{1}{\sqrt{2}}(|HHH\rangle\pm|VVV\rangle)_{ABC},\nonumber\\
|\Phi_{1}^{\pm}\rangle_{ABC}=\frac{1}{\sqrt{2}}(|VHH\rangle\pm|HVV\rangle)_{ABC},\nonumber\\
|\Phi_{2}^{\pm}\rangle_{ABC}=\frac{1}{\sqrt{2}}(|HVH\rangle\pm|VHV\rangle)_{ABC},\nonumber\\
|\Phi_{3}^{\pm}\rangle_{ABC}=\frac{1}{\sqrt{2}}(|HHV\rangle\pm|VVH\rangle)_{ABC}.\label{GHZstate}
\end{eqnarray}
Here the subscripts $A$, $B$, and  $C$ represent the photons
(qubits) sent to the parties Alice, Bob, and Charlie, respectively.
Suppose that the original GHZ state transmitted among the three
parties is $|\Phi_{0}^{+}\rangle_{ABC}$. If a bit-flip error takes
place on the first qubit in this GHZ state after it is transmitted
over a noisy channel, the three-photon system is in the state
$|\Phi^{+}_{1}\rangle_{ABC}$. $|\Phi^{+}_{2}\rangle_{ABC}$ and
$|\Phi^{+}_{3}\rangle_{ABC}$ represent the instances that a bit-flip
error takes place on the second qubit and the third qubit,
respectively. If $|\Phi_{0}^{+}\rangle$ becomes
$|\Phi_{0}^{-}\rangle$, there is a phase-flip error. Sometimes, both
a bit-flip error and a phase-flip error will take place on a
three-photon quantum system transmitted over a noisy channel. The
task for purifying three-photon entangled systems requires to
correct both bit-flip errors and phase-flip errors on the quantum
system. We first discuss the principle of the present MEPP for
purifying the bit-flip errors and discuss it for phase-flip errors
in next section.

Suppose that Alice, Bob and Charlie share a three-qubit ensemble
$\rho$ after the transmission of qubits over  noisy channels, that
is,
\begin{eqnarray}
\rho &=& F_0|\Phi_{0}^{+}\rangle\langle\Phi_{0}^{+}| +
F_1|\Phi_{1}^{+}\rangle\langle\Phi_{1}^{+}|   \nonumber\\
&+& F_2|\Phi_{2}^{+}\rangle\langle\Phi_{2}^{+}| +
F_3|\Phi_{3}^{+}\rangle\langle\Phi_{3}^{+}|.\label{ensemblerho}
\end{eqnarray}
Here  $F_0=\langle \Phi_{0}^+|\rho|\Phi_{0}^+\rangle$  is the
fidelity of the quantum systems transmitted over noisy channels, and
\begin{eqnarray}
F_0+F_1+F_2+F_3=1.
\end{eqnarray}
The density matrix $\rho$  means that there is a bit-flip error on
the first qubit, the second qubit, and the third qubit of the
quantum system with a probability of $F_1$, $F_2$, and $F_3$,
respectively.  For obtaining some high-fidelity entangled
three-photon  systems, the three parties divide their quantum
systems in the ensemble $\rho$ into many groups and each group is
composed of a pair of three-photon quantum systems. We label each
group with $A_1B_1C_1A_2B_2C_2$ (that is,  the two three-photon
quantum systems $A_1B_1C_1$ and $A_2B_2C_2$).

\begin{figure}[!h]
\begin{center}
\includegraphics[width=8cm,angle=0]{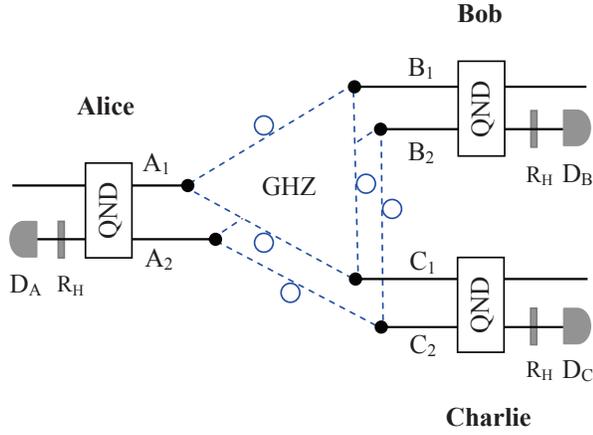}
\caption{(Color online) The principle of our conventional
three-photon entanglement purification scheme for bit-flip errors
with QNDs. The wave plate $R_{H}$ represents a Hadamard operation
and it is used to transform the polarization states $|H\rangle$  and
$|V\rangle$ into $1/\sqrt{2}(|H\rangle+|V\rangle)$ and
$1/\sqrt{2}(|H\rangle-|V\rangle)$, respectively. $D_A$,  $D_B$, and
$D_C$ represent the single-photon measurements with the basis
$Z=\{\vert H\rangle, \vert V\rangle\}$ done by Alice, Bob, and
Charlie, respectively. The circles with blue virtual lines represent
long-distance optical-fiber channels. The dots with black real lines
represent qubits.} \label{fig2_three_qubit}
\end{center}
\end{figure}

The principle of our conventional three-photon EPP for  bit-flip
errors is shown in Fig.\ref{fig2_three_qubit}. The state of the
system composed of the two three-photon subsystems $A_1B_1C_1$ and
$A_2B_2C_2$ can be viewed as the mixture of the sixteen pure states,
that is,  $|\Phi_{i}^{+}\rangle\otimes|\Phi_{j}^{+}\rangle$ with a
probability of $F_iF_j$ ($i,j=0,1,2,3$). For each group, Alice takes
her two photons $A_1$ and $A_2$ to pass through the setup shown in
Fig.\ref{fig2_three_qubit}. The photon $A_1$ enters the up spatial
mode and the photon $A_2$ enters the down spatial mode. So do Bob
and Charlie. After the QNDs, the three parties compare the parities
of their photons.  They  keep the groups for which all the three
parties obtain an even parity or an odd parity. These instances
correspond to the identity-combinations $\vert
\Phi^+_i\rangle_{A_1B_1C_1} \otimes \vert
\Phi^+_i\rangle_{A_2B_2C_2}$ ($i=0,1,2,3$). That is, the parties
only distill some high-fidelity three-photon entangled systems from
the identity-combinations in our conventional MEPP, similar to all
existing EPPs  \cite{
Bennett1996,Deutsch,Pan1,Simon,shengpra2008,Murao,shengepjd,shengpla}.

When all the three parties obtain an even parity, the quantum system
$A_1B_1C_1A_2B_2C_2$ is in a new mixed state which is composed of
the four states
\begin{eqnarray}
|\phi_0\rangle &=&
\frac{1}{\sqrt{2}}(|HHH\rangle_{A_1B_1C_1}|HHH\rangle_{A_2B_2C_2}
\nonumber\\
&& \;\;\; +\; |VVV\rangle_{A_1B_1C_1}|VVV\rangle_{A_2B_2C_2}),
\end{eqnarray}
\begin{eqnarray}
|\phi_{1}\rangle&=&
\frac{1}{\sqrt{2}}(|VHH\rangle_{A_1B_1C_1}|VHH\rangle_{A_2B_2C_2}
\nonumber\\
&& \;\;\; +\; |HVV\rangle_{A_1B_1C_1}|HVV\rangle_{A_2B_2C_2}),
\end{eqnarray}
\begin{eqnarray}
|\phi_{2}\rangle&=&
\frac{1}{\sqrt{2}}(|HVH\rangle_{A_1B_1C_1}|HVH\rangle_{A_2B_2C_2}
\nonumber\\
&& \;\;\; +\; |VHV\rangle_{A_1B_1C_1}|VHV\rangle_{A_2B_2C_2}),
\end{eqnarray}
\begin{eqnarray}
|\phi_{3}\rangle&=&
\frac{1}{\sqrt{2}}(|HHV\rangle_{A_1B_1C_1}|HHV\rangle_{A_2B_2C_2}
\nonumber\\
&& \;\;\; +\; |VVH\rangle_{A_1B_1C_1}|VVH\rangle_{A_2B_2C_2})
\end{eqnarray}
with a probability of $\frac{1}{2} F_{0}^{2}$,  $\frac{1}{2}F^2_1$,
$\frac{1}{2}F^2_2$, and $\frac{1}{2}F_{3}^2$, respectively.

When all the three parties obtain an odd parity, the quantum system
$A_1B_1C_1A_2B_2C_2$ is in another mixed state which is composed of
the four states
\begin{eqnarray}
|\psi_0\rangle &=&
\frac{1}{\sqrt{2}}(|HHH\rangle_{A_1B_1C_1}|VVV\rangle_{A_2B_2C_2}
\nonumber\\
&& \;\;\; +\; |VVV\rangle_{A_1B_1C_1}|HHH\rangle_{A_2B_2C_2}),
\end{eqnarray}
\begin{eqnarray}
|\psi_{1}\rangle&=&
\frac{1}{\sqrt{2}}(|VHH\rangle_{A_1B_1C_1}|HVV\rangle_{A_2B_2C_2}
\nonumber\\
&& \;\;\; +\; |HVV\rangle_{A_1B_1C_1}|VHH\rangle_{A_2B_2C_2}),
\end{eqnarray}
\begin{eqnarray}
|\psi_{2}\rangle&=&
\frac{1}{\sqrt{2}}(|HVH\rangle_{A_1B_1C_1}|VHV\rangle_{A_2B_2C_2}
\nonumber\\
&& \;\;\; +\; |VHV\rangle_{A_1B_1C_1}|HVH\rangle_{A_2B_2C_2}),
\end{eqnarray}
\begin{eqnarray}
|\psi_{3}\rangle&=&
\frac{1}{\sqrt{2}}(|HHV\rangle_{A_1B_1C_1}|VVH\rangle_{A_2B_2C_2}
\nonumber\\
&& \;\;\; +\; |VVH\rangle_{A_1B_1C_1}|HHV\rangle_{A_2B_2C_2})
\end{eqnarray}
with the probabilities of $\frac{1}{2} F_{0}^{2}$,
$\frac{1}{2}F^2_1$, $\frac{1}{2}F^2_2$, and $\frac{1}{2}F_{3}^2$,
respectively. With a bit-flip operation on each of the three qubits
$A_2B_2C_2$, Alice, Bob, and Charlie obtain the same outcomes as the
case in which they all obtain an even parity. That is, the states
$\vert \psi_i\rangle$ can be transformed into the states $\vert
\phi_i\rangle$ ($i=0,1,2,3$), respectively. By this way,  Alice,
Bob, and Charlie obtain the states $\vert \phi_i\rangle$ with the
probabilities $F_i^2$ as they can obtain the similar outcomes when
they all get whether an even parity or an odd parity with QNDs. We
only discuss the case that the system is in the states $\vert
\phi_i\rangle$ with the probabilities $F_i^2$  below.

After a Hadamard operation (that is,  the wave plate $R_H$ in
Fig.\ref{fig2_three_qubit}) on the polarization of each photon in
the down spatial mode, Alice, Bob, and Charlie measure the photons
$A_2B_2C_2$ with the basis $\{\vert H\rangle, \vert V\rangle\}$. The
outcomes will divide the instances into two groups. In the first
group, the number of the outcomes $\vert V\rangle$ is even. In this
time, Alice, Bob, and Charlie obtain the states $\vert
\Phi_{0}^+\rangle_{A_1B_1C_1}$, $\vert
\Phi_{1}^+\rangle_{A_1B_1C_1}$, $\vert
\Phi_{2}^+\rangle_{A_1B_1C_1}$, and $\vert
\Phi_{3}^+\rangle_{A_1B_1C_1}$ with the probabilities
$\frac{1}{2}F_{0}^2$, $\frac{1}{2}F_{1}^2$, $\frac{1}{2}F_{2}^2$,
and $\frac{1}{2}F_{3}^2$, respectively. In the second group, the
number of the outcomes $\vert V\rangle$  is odd and the three
parties obtain the states $\vert \Phi_{0}^-\rangle_{A_1B_1C_1}$,
$\vert \Phi_{1}^-\rangle_{A_1B_1C_1}$, $\vert
\Phi_{2}^-\rangle_{A_1B_1C_1}$, and $\vert
\Phi_{3}^-\rangle_{A_1B_1C_1}$ with the probabilities
$\frac{1}{2}F_{0}^2$, $\frac{1}{2}F_{1}^2$, $\frac{1}{2}F_{2}^2$,
and $\frac{1}{2}F_{3}^2$, respectively. Alice, Bob, and Charlie can
transform the states  $\vert \Phi_{0}^-\rangle$,   $\vert
\Phi_{1}^-\rangle$,   $\vert \Phi_{2}^-\rangle$, and  $\vert
\Phi_{3}^-\rangle$ into the states $\vert \Phi_{0}^+\rangle$, $\vert
\Phi_{1}^+\rangle$,   $\vert \Phi_{2}^+\rangle$, and  $\vert
\Phi_{3}^+\rangle$ with a phase-flip operation
$\sigma_z=|H\rangle\langle H| - |V\rangle\langle V|$ on the first
photon $A_1$, respectively. That is, by keeping the instances in
which all the three parties obtain the same parity and measuring the
photons through the down spatial modes, Alice, Bob, and Charlie can
obtain a new ensemble in the state
\begin{eqnarray}
\rho'_3 &=& F'_{0} |\Phi_{0}^{+}\rangle\langle\Phi_{0}^{+}| +
F'_{1}|\Phi_{1}^{+}\rangle\langle\Phi_{1}^{+}| \nonumber\\
& + & F'_{2}|\Phi_{2}^{+}\rangle\langle\Phi_{2}^{+}| +
F'_{3}|\Phi_{3}^{+}\rangle\langle\Phi_{3}^{+}|,
\end{eqnarray}
where
\begin{eqnarray}
F'_0 &=& \frac{F_{0}^2}{F_{0}^2 + F_{1}^2 + F_{2}^2 +
(1-F_{0}-F_{1}-F_{2})^2},\nonumber\\
F'_1 &=& \frac{F_{1}^2}{F_{0}^2 + F_{1}^2 + F_{2}^2 +
(1-F_{0}-F_{1}-F_{2})^2},\nonumber\\
F'_2 &=& \frac{F_{2}^2}{F_{0}^2 + F_{1}^2 + F_{2}^2 +
(1-F_{0}-F_{1}-F_{2})^2},\nonumber\\
F'_3 &=& \frac{(1-F_{0}-F_{1}-F_{2})^2}{F_{0}^2 + F_{1}^2 + F_{2}^2
+ (1-F_{0}-F_{1}-F_{2})^2}.\label{conventionalfidelity3}
\end{eqnarray}
The fidelity  $F'_0>F_0$ if $F_0$ satisfies the relation
\begin{eqnarray}
F_0 &>& \frac{1}{4}\{3-2F_1-2F_2 \nonumber\\
&& -\sqrt{1+4(F_1+F_2)-12(F_{1}^2+F_{2}^2)-8F_1F_2}\}.\nonumber\\
\end{eqnarray}
With three symmetric noisy channels, the fidelity of the state
$\vert \Phi_{0}^+\rangle$ will be improved by this conventional
three-photon EPP if its original fidelity $F_0>1/4$.

In fact, our conventional three-photon EPP is similar to the MEPP in
Ref. \cite{Murao}. We use some QNDs, instead of perfect CNOT gates
in Ref. \cite{Murao}, to complete the purification of bit-flip
errors and we give a general case for this purification.

\subsection{Recycling three-photon entanglement purification for bit-flip errors from subspaces}

In our conventional three-photon entanglement purification for
bit-flip errors, the three parties in quantum communication discard
the instances in which Alice, Bob, and Charlie obtain different
parities, that is,  the cross-combinations $\vert\Phi_i^+\rangle
\otimes \vert\Phi_j^+\rangle$ ($i\neq j \in\{0,1,2,3\}$), similar to
all conventional EPPs \cite{
Bennett1996,Deutsch,Pan1,Simon,shengpra2008,Murao,shengepjd,shengpla}.
That is, when the system composed of six photons is in the state
$\vert\Phi_i^+\rangle \otimes \vert\Phi_j^+\rangle$ ($i \neq j
\in\{0,1,2,3\}$) which take place with a  probability of $F_iF_j$,
Alice, Bob, and Charlie discard the system in the conventional
three-photon EPPs because the probabilities of the states
$\vert\Phi_i^+\rangle_{A_1B_1C_1} \otimes
\vert\Phi_j^+\rangle_{A_2B_2C_2} $ and
$\vert\Phi_j^+\rangle_{A_1B_1C_1}  \otimes
\vert\Phi_i^+\rangle_{A_2B_2C_2}$ are the same one  $F_iF_j$ and the
three parties cannot determine the state of the three-photon system
$A_1B_1C_1$ after they  measure the photons $A_2B_2C_2$ in this
time. However, this system can be used to distil a high-fidelity
two-photon entangled state. With a set of high-fidelity two-photon
systems, Alice, Bob, and Charlie can produce a subset of
high-fidelity three-photon systems. We call this part of our MEPP as
the recycling MEPP as the parties should distil three-photon systems
from the cross-combinations $\vert\Phi_i^+\rangle \otimes
\vert\Phi_j^+\rangle$ ($i\neq j \in\{0,1,2,3\}$)  which are just
discarded in all other conventional MEPPs. Our recycling MEPP will
increase the efficiency (the yield) of our three-photon EPP largely.

\subsubsection{Two-photon entanglement purification for bit-flip
errors from three-photon systems}
 \label{seciic}

We only discuss the principle of our recycling MEPP in the case that
the system  is in the state
$|\Phi_{0}^{+}\rangle_{A_1B_1C_1}\otimes|\Phi^{+}_{2}\rangle_{A_2B_2C_2}$
or
$|\Phi^{+}_{2}\rangle_{A_1B_1C_1}\otimes|\Phi_{0}^{+}\rangle_{A_2B_2C_2}$
below and the other cases are similar to it with or without a little
modification.

The cross-combinations $|\Pi\rangle_1 \equiv
|\Phi_0^{+}\rangle_{A_1B_1C_1}\otimes|\Phi^{+}_{2}\rangle_{A_2B_2C_2}$
and $|\Pi\rangle_2 \equiv
|\Phi^{+}_{2}\rangle_{A_1B_1C_1}\otimes|\Phi_0^{+}\rangle_{A_2B_2C_2}$
can be rewritten as
\begin{eqnarray}
|\Pi\rangle_1
&  =&\frac{1}{2}(|HHH\rangle_{A_1B_1C_1}|VHV\rangle_{A_2B_2C_2} \nonumber\\
&&\;   + \;  |VVV\rangle_{A_1B_1C_1}|HVH\rangle_{A_2B_2C_2}  \nonumber\\
&&\;   + \; |HHH\rangle_{A_1B_1C_1}|HVH\rangle_{A_2B_2C_2} \nonumber\\
&&\;   + \;  |VVV\rangle_{A_1B_1C_1}|VHV\rangle_{A_2B_2C_2}),\\
|\Pi\rangle_2
&  =& \frac{1}{2}(|VHV\rangle_{A_1B_1C_1}|HHH\rangle_{A_2B_2C_2} \nonumber\\
&&\;   + \;  |HVH\rangle_{A_1B_1C_1}|VVV\rangle_{A_2B_2C_2}  \nonumber\\
&&\;   + \;  |HVH\rangle_{A_1B_1C_1}|HHH\rangle_{A_2B_2C_2} \nonumber\\
&&\;   + \;  |VHV\rangle_{A_1B_1C_1}|VVV\rangle_{A_2B_2C_2}).
\end{eqnarray}
That is, if the outcomes of the parity-check measurements obtained
by Alice, Bob, and Charlie are odd, even, and odd, respectively, the
six-photon system is in the state
\begin{eqnarray}
|\Omega\rangle_1 &\equiv&
\frac{1}{\sqrt{2}}(|HHH\rangle_{A_1B_1C_1}
|VHV\rangle_{A_2B_2C_2}\nonumber\\
& + & |VVV\rangle_{A_1B_1C_1} |HVH\rangle_{A_2B_2C_2})
\end{eqnarray}
 or
\begin{eqnarray}
|\Omega\rangle_2 &\equiv& \frac{1}{\sqrt{2}}(|VHV\rangle_{A_1B_1C_1}
|HHH\rangle_{A_2B_2C_2} \nonumber\\
&+& |HVH\rangle_{A_1B_1C_1} |VVV\rangle_{A_2B_2C_2}),
\end{eqnarray}
 which takes place with the probability of
$F_0F_1$. If they are even, odd, and even, respectively, the
six-photon system is in the state
\begin{eqnarray}
|\Omega\rangle_3 &\equiv&
\frac{1}{\sqrt{2}}(|HHH\rangle_{A_1B_1C_1}|HVH\rangle_{A_2B_2C_2} \nonumber\\
&+& |VVV\rangle_{A_1B_1C_1}|VHV\rangle_{A_2B_2C_2})
\end{eqnarray}
 or
\begin{eqnarray}
|\Omega\rangle_4 &\equiv&
\frac{1}{\sqrt{2}}(|HVH\rangle_{A_1B_1C_1}|HHH\rangle_{A_2B_2C_2} \nonumber\\
&+& |VHV\rangle_{A_1B_1C_1}|VVV\rangle_{A_2B_2C_2}),
\end{eqnarray}
which takes place with the probability of $F_0F_1$ yet.

The states $|\Omega\rangle_1$ can be rewritten as follows:
\begin{eqnarray}
|\Omega\rangle_1 &=& \frac{1}{2\sqrt{2}}\{(|HH\rangle_{A_1C_1} +
|VV\rangle_{A_1C_1})
(\vert ++++\rangle  \nonumber \\
&&\;\;\;\;   + \;  \vert ++--\rangle - \vert +-+-\rangle \nonumber \\
&&\;\;\;\; - \; \vert +--+\rangle - \vert -++-\rangle \nonumber \\
&&\;\;\;\; - \; \vert -+-+\rangle  + \vert
--++\rangle \nonumber \\
&&\;\;\;\;   + \; \vert ----\rangle)_{A_2C_2B_1B_2}\nonumber \\
&&\;\;\;\;  + \; (|HH\rangle_{A_1C_1} - |VV\rangle_{A_1C_1})
(\vert +++-\rangle \nonumber \\
&&\;\;\;\;   + \;  \vert ++-+\rangle - \vert +-++\rangle
\nonumber \\
&&\;\;\;\; - \; \vert +---\rangle - \vert -+++\rangle \nonumber \\
&&\;\;\;\; - \; \vert -+--\rangle  + \vert
--+-\rangle \nonumber \\
&&\;\;\;\;   + \; \vert ---+\rangle)_{A_2C_2B_1B_2}\},
\end{eqnarray}
where
\begin{eqnarray}
|+\rangle  &=&  \frac{1}{\sqrt{2}}(|H\rangle + |V\rangle),\nonumber\\
|-\rangle  &=&  \frac{1}{\sqrt{2}}(|H\rangle - |V\rangle).
\end{eqnarray}
That is, Alice, Bob, and Charlie can distil a high-fidelity
two-photon entangled state  $\vert
\phi^+\rangle_{A_1C_1}=\frac{1}{\sqrt{2}}(\vert HH\rangle + \vert
VV\rangle)_{A_1C_1}$ from  the six-photon state $|\Omega\rangle_1$.
In detail, Alice and Charlie measure their photons $A_2$ and $C_2$,
respectively, and Bob measures his two photons $B_1$ and $B_2$ with
the measuring basis $X\equiv \{\vert +\rangle, \vert -\rangle\}$.
Alice and Charlie   obtain the two-photon entangled state $\vert
\phi^+\rangle_{A_1C_1}$ from the six-photon state $|\Omega\rangle_1$
when the number of the outcomes $\vert -\rangle$ is even. When the
number of the outcomes $\vert -\rangle$ is odd, Alice and Charlie
 obtain the two-photon entangled state $\vert
\phi^-\rangle_{A_1C_1}=\frac{1}{\sqrt{2}}(\vert HH\rangle - \vert
VV\rangle)_{A_1C_1}$ and they can transform the state $\vert
\phi^-\rangle_{A_1C_1}$ into the state $\vert
\phi^+\rangle_{A_1C_1}$ by performing a phase-flip operation
$\sigma_z$ on the photon $C_1$.

For the other three states $|\Omega\rangle_i$ ($i=2,3,4$), Alice,
Bob, and Charlie can also obtain the two-photon entangled state
$\vert \phi^+\rangle_{A_1C_1}$ with the same principle. That is,
Alice, Bob, and Charlie can obtain the two-photon maximally
entangled state $\vert
\phi^+\rangle_{A_1C_1}=\frac{1}{\sqrt{2}}(\vert HH\rangle + \vert
VV\rangle)_{A_1C_1}$ from the states
$|\Phi_{0}^{+}\rangle_{A_1B_1C_1}\otimes|\Phi^{+}_{2}\rangle_{A_2B_2C_2}$
and
$|\Phi^{+}_{2}\rangle_{A_1B_1C_1}\otimes|\Phi_{0}^{+}\rangle_{A_2B_2C_2}$
with the probability of $2F_0F_2$.

In the same way, Alice, Bob, and Charlie can obtain the two-photon
maximally entangled states $\vert
\phi^+\rangle_{A_1B_1}=\frac{1}{\sqrt{2}}(\vert HH\rangle + \vert
VV\rangle)_{A_1B_1}$ from the cross-combinations
$|\Phi_{0}^{+}\rangle_{A_1B_1C_1}\otimes|\Phi^{+}_{3}\rangle_{A_2B_2C_2}$
and
$|\Phi^{+}_{3}\rangle_{A_1B_1C_1}\otimes|\Phi_{0}^{+}\rangle_{A_2B_2C_2}$
 with the probability of  $2F_0F_3$. Also, they can obtain the
$\vert \phi^+\rangle_{B_1C_1}=\frac{1}{\sqrt{2}}(\vert HH\rangle +
\vert VV\rangle)_{B_1C_1}$ from
$|\Phi_{0}^{+}\rangle_{A_1B_1C_1}\otimes|\Phi^{+}_{1}\rangle_{A_2B_2C_2}$
and
$|\Phi^{+}_{1}\rangle_{A_1B_1C_1}\otimes|\Phi_{0}^{+}\rangle_{A_2B_2C_2}$
with the probability of $2F_0F_1$.

Certainly, there is a probability that the two three-photon systems
take place a bit-flip error on two different photons, such as the
states
$|\Phi_{1}^{+}\rangle_{A_1B_1C_1}\otimes|\Phi^{+}_{2}\rangle_{A_2B_2C_2}$,
$|\Phi_{2}^{+}\rangle_{A_1B_1C_1}\otimes|\Phi^{+}_{1}\rangle_{A_2B_2C_2}$,
$|\Phi_{1}^{+}\rangle_{A_1B_1C_1}\otimes|\Phi^{+}_{3}\rangle_{A_2B_2C_2}$,
$|\Phi_{3}^{+}\rangle_{A_1B_1C_1}\otimes|\Phi^{+}_{1}\rangle_{A_2B_2C_2}$,
$|\Phi_{2}^{+}\rangle_{A_1B_1C_1}\otimes|\Phi^{+}_{3}\rangle_{A_2B_2C_2}$,
and
$|\Phi_{3}^{+}\rangle_{A_1B_1C_1}\otimes|\Phi^{+}_{2}\rangle_{A_2B_2C_2}$.
With the same process as the case that there is only one
three-photon system taking place a bit-flip error (that is,  the
states
$|\Phi_{0}^{+}\rangle_{A_1B_1C_1}\otimes|\Phi^{+}_{i}\rangle_{A_2B_2C_2}$
and
$|\Phi_{i}^{+}\rangle_{A_1B_1C_1}\otimes|\Phi^{+}_{0}\rangle_{A_2B_2C_2}$
($i=1,2,3$)), Alice, Bob, and Charlie will obtain the states with a
bit-flip error, such as $\vert
\psi^+\rangle_{A_1B_1}=\frac{1}{\sqrt{2}}(\vert HV\rangle + \vert
VH\rangle)_{A_1B_1}$,  $\vert
\psi^+\rangle_{A_1C_1}=\frac{1}{\sqrt{2}}(\vert HV\rangle + \vert
VH\rangle)_{A_1C_1}$, and  $\vert
\psi^+\rangle_{B_1C_1}=\frac{1}{\sqrt{2}}(\vert HV\rangle + \vert
VH\rangle)_{B_1C_1}$ with the probabilities $2F_1F_2$, $2F_1F_3$,
and $2F_2F_3$, respectively. The relation between the states of the
two-photon systems and the cross-combinations is shown in Table
\ref{tab1}.

\begin{widetext}
\begin{center}
\begin{table}[!h]
\tabcolsep 0pt \caption{The states of the two-photon systems
obtained from cross-combinations and their probabilities. } 
\begin{center}
\begin{tabular}{c|cccccc}\hline
      cross-                & \;\;\;\; $\vert \Phi^+_0\rangle\otimes \vert \Phi^+_2\rangle \;\;\;\;$   &   $\vert \Phi^+_0\rangle\otimes \vert \Phi^+_1\rangle  \;\;\;\;$   &   $\vert \Phi^+_0\rangle\otimes \vert \Phi^+_3\rangle  \;\;\;\;$  &   $\vert \Phi^+_1\rangle\otimes \vert \Phi^+_2\rangle  \;\;\;\;$  &    $\vert \Phi^+_1\rangle\otimes \vert \Phi^+_3\rangle  \;\;\;\;$  &   $\vert \Phi^+_2\rangle\otimes \vert \Phi^+_3\rangle $      \\
     combinations           & \;\;\;\; $\vert \Phi^+_2\rangle\otimes \vert \Phi^+_0\rangle \;\;\;\;$   &   $\vert \Phi^+_1\rangle\otimes \vert \Phi^+_0\rangle \;\;\;\;$    &   $\vert \Phi^+_3\rangle\otimes \vert \Phi^+_0\rangle \;\;\;\;$   &   $\vert \Phi^+_2\rangle\otimes \vert \Phi^+_1\rangle \;\;\;\;$   &    $\vert \Phi^+_3\rangle\otimes \vert \Phi^+_1\rangle \;\;\;\;$   &   $\vert \Phi^+_3\rangle\otimes \vert \Phi^+_2\rangle$           \\\hline
       two-photon states    & \;\;\;\; $\vert \phi^+\rangle_{A_1C_1} \;\;\;\;$                         &   $\vert \phi^+\rangle_{B_1C_1} \;\;\;\;$                          &   $\vert \phi^+\rangle_{A_1B_1} \;\;\;\;$                         &   $\vert \psi^+\rangle_{A_1B_1} \;\;\;\;$                         &    $\vert \psi^+\rangle_{A_1C_1} \;\;\;\;$                         &   $\vert \psi^+\rangle_{B_1C_1}$                              \\\hline
       probabilities        & \;\;\;\; $2F_0F_2 \;\;\;\;$                                              &   $2F_0F_1 \;\;\;\;$                                               &   $2F_0F_3 \;\;\;\;$                                              &   $2F_1F_2 \;\;\;\;$                                              &    $2F_1F_3 \;\;\;\;$                                              &   $2F_2F_3$
\\\hline
\end{tabular}\label{tab1}
       \end{center}
       \end{table}
       \end{center}
\end{widetext}

With Table \ref{tab1}, one can see that the two-photon systems
shared by two of the three parties can be described with the
following density matrices (without normalization):
\begin{eqnarray}
\rho_{AB}  &=&  2F_0F_3|\phi^+\rangle_{AB}\langle \phi^+\vert + 2F_1F_2|\psi^+\rangle_{AB}\langle \psi^+\vert, \nonumber\\
\rho_{AC}  &=&  2F_0F_2|\phi^+\rangle_{AC}\langle \phi^+\vert + 2F_1F_3|\psi^+\rangle_{AC}\langle \psi^+\vert, \nonumber\\
\rho_{BC}  &=&  2F_0F_1|\phi^+\rangle_{BC}\langle \phi^+\vert +
2F_2F_3|\psi^+\rangle_{BC}\langle \psi^+\vert.\label{dmatrix1}
\end{eqnarray}
Suppose the fidelities $F_1=F_2=F_3$ and $F_0>F_1$. One can see that
the fidelity of two-photon systems is larger than that of the
original three-photon systems transmitted. For example,
$F(\vert\phi^+\rangle_{AB} )=\frac{2F_0F_3}{2F_0F_3 +
2F_1F_2}=\frac{F_0}{F_0 + F_1}>F_0$ as $F_0+F_1<1$.

\begin{figure}[!h]
\begin{center}
\includegraphics[width=6.4cm,angle=0]{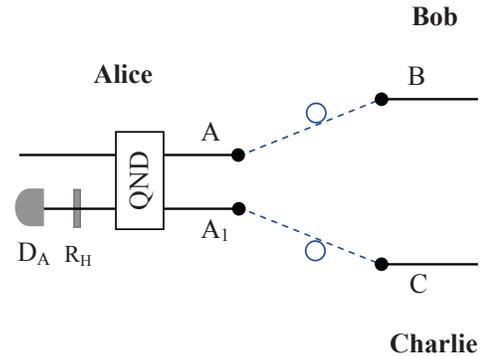}
\caption{(Color online)  The principle of the entanglement link for
producing a three-photon entangled system from two two-photon
entangled systems with a QND. } \label{fig3_two_qubit}
\end{center}
\end{figure}

\subsubsection{Three-photon entanglement production from two-photon
systems with entanglement link} \label{seciid}

As the three photons in the original system are symmetric to each
other, we use the states $\rho_{AB}$ and $\rho_{AC}$ as an example
to describe the principle of three-photon entanglement production
from two-photon systems with entanglement link and assume that
$F_1=F_2=F_3$, shown  in Fig.\ref{fig3_two_qubit}. The density
matrices in Eq.(\ref{dmatrix1}) become
\begin{eqnarray}
\rho^b_{AB}  &=&  F^b_0 |\phi^+\rangle_{AB}\langle \phi^+\vert +  F^b_1  |\psi^+\rangle_{AB}\langle \psi^+\vert, \nonumber\\
\rho^b_{AC}  &=&  F^b_0 |\phi^+\rangle_{AC}\langle \phi^+\vert +  F^b_1  |\psi^+\rangle_{AC}\langle \psi^+\vert, \nonumber\\
\rho^b_{BC}  &=&  F^b_0 |\phi^+\rangle_{BC}\langle \phi^+\vert +  F^b_1  |\psi^+\rangle_{BC}\langle \psi^+\vert,
\end{eqnarray}
where
\begin{eqnarray}
F^b_0 &=& \frac{F_0}{F_0 + F_1}, \nonumber\\
F^b_1 &=& \frac{F_1}{F_0 + F_1}.
\end{eqnarray}
The system composed of the four photons $A$, $B$, $A_1$, and $C$ is
in the state $\rho^b_{AB}\otimes \rho^b_{A_1C}$ which can be viewed
as the mixture of the four pure states  $\vert \Gamma\rangle_0
\equiv \vert \phi^+\rangle_{AB} \otimes \vert \phi^+\rangle_{A_1C}$,
$\vert \Gamma\rangle_1 \equiv \vert \phi^+\rangle_{AB} \otimes \vert
\psi^+\rangle_{A_1C}$, $\vert \Gamma\rangle_2 \equiv \vert
\psi^+\rangle_{AB} \otimes \vert \phi^+\rangle_{A_1C}$, and $\vert
\Gamma\rangle_3 \equiv \vert \psi^+\rangle_{AB} \otimes \vert
\psi^+\rangle_{A_1C}$ with the probabilities $F^b_0F^b_0$,
$F^b_0F^b_1$,  $F^b_1F^b_0$, and $F^b_1F^b_1$, respectively. After
the QND, Alice, Bob, and Charlie will divide the instances into two
groups. One is the case that the outcome of the parity-check
measurement is even and the other is odd. When Alice obtains an even
parity, the four-photon system is in the states
$\frac{1}{\sqrt{2}}(\vert HHHH\rangle + \vert
VVVV\rangle)_{ABA_1C}$, $\frac{1}{\sqrt{2}}(\vert HHHV\rangle +
\vert VVVH\rangle)_{ABA_1C}$, $\frac{1}{\sqrt{2}}(\vert HVHH\rangle
+ \vert VHVV\rangle)_{ABA_1C}$, and $\frac{1}{\sqrt{2}}(\vert
HVHV\rangle + \vert VHVH\rangle)_{ABA_1C}$ with the probabilities
$\frac{1}{2}F^s_0F^s_0$, $\frac{1}{2}F^b_0F^b_1$,
$\frac{1}{2}F^b_1F^b_0$, and $\frac{1}{2}F^b_1F^b_1$, respectively.
After Alice performs a Hadamard operation on the photon $A_1$ and
measures it with the basis $Z=\{\vert H\rangle, \vert V\rangle\}$,
Alice, Bob, and Charlie will obtain a three-photon entangled system
in the states $\vert\Phi^+_0\rangle$, $\vert\Phi^+_3\rangle$,
$\vert\Phi^+_2\rangle$, and $\vert\Phi^+_1\rangle$ with the
probabilities $\frac{1}{2}F^b_0F^b_0$, $\frac{1}{2}F^b_0F^b_1$,
$\frac{1}{2}F^b_1F^b_0$, and $\frac{1}{2}F^b_1F^b_1$, respectively.
These outcomes will be obtained with (if the outcome of the
measurement on the photon $A_1$ by Alice is $\vert V\rangle_{A_1}$)
or without ($\vert H\rangle_{A_1}$) a phase-flip operation on the
photon $A$.

When Alice obtains an odd parity with her QND, the four-photon
system is in the states $\frac{1}{\sqrt{2}}(\vert HHVV\rangle +
\vert VVHH\rangle)_{ABA_1C}$, $\frac{1}{\sqrt{2}}(\vert HHVH\rangle
+ \vert VVHV\rangle)_{ABA_1C}$, $\frac{1}{\sqrt{2}}(\vert
HVVV\rangle + \vert VHHH\rangle)_{ABA_1C}$, and
$\frac{1}{\sqrt{2}}(\vert HVVH\rangle + \vert VHHV\rangle)_{ABA_1C}$
with the probabilities $\frac{1}{2}F^b_0F^b_0$,
$\frac{1}{2}F^b_0F^b_1$, $\frac{1}{2}F^b_1F^b_0$, and
$\frac{1}{2}F^b_1F^b_1$, respectively. Alice, Bob, and Charlie can
obtain the same outcomes as the case with an even parity by
performing a bit-flip operation on the photons $A_1$ and $C$
independently. That is, with entanglement link, Alice, Bob, and
Charlie can obtain a new ensemble for three-photon systems in the
state
\begin{eqnarray}
\rho^t &=& F^t_0|\Phi_{0}^{+}\rangle\langle\Phi_{0}^{+}| +
F^t_1|\Phi_{1}^{+}\rangle\langle\Phi_{1}^{+}|   \nonumber\\
&+& F^t_2|\Phi_{2}^{+}\rangle\langle\Phi_{2}^{+}| +
F^t_3|\Phi_{3}^{+}\rangle\langle\Phi_{3}^{+}|.\label{ensemblerho}
\end{eqnarray}
Here
\begin{eqnarray}
F^t_0 &=& \frac{F^2_0}{(F_0+F_1)^2 }, \nonumber\\
F^t_1 &=& \frac{F^2_1}{(F_0+F_1)^2 }, \nonumber\\
F^t_2 &=& F^t_3 = \frac{F_0F_1}{(F_0+F_1)^2 }.
\end{eqnarray}
$F^t_0>F_0$ when $F_0>\frac{1}{4}$, which means that the three
parties can obtain a high-fidelity three-photon entangled system
from two two-photon entangled subsystems if and only if the original
fidelity of the three-photon systems transmitted over noisy channels
is larger than $\frac{1}{4}$ (this is just the condition for the
conventional three-photon entanglement purification over three
symmetric noisy channels).

\subsection{Efficiency and fidelity of the present three-photon entanglement purification protocol}

\label{seciie}

Here the efficiency of an EPP $Y$ is defined as the probability
(that is,  the yield) that the parties can obtain a high-fidelity
entangled multi-photon system from a pair of multi-photon systems
transmitted over a noisy channel without loss. The efficiency of the
present three-photon EPP $Y_e$ depends on the parameters $F_1$,
$F_2$, and $F_3$. For simpleness, we only discuss the case with the
parameters $F_1=F_2=F_3=\frac{1-F_0}{3}$ below.

With our conventional three-photon EPP only, the efficiency of the
three-photon EPP $Y_c$ is
\begin{eqnarray}
Y_c &=& F^2_0+F^2_1+F^2_2+F^2_3
= \frac{1 - 2F_0 + 4F^2_0}{3}.
\end{eqnarray}
It is the probability that the pair of three-photon systems are in
the identity-combinations $\vert \Phi^+_i\rangle_{A_1B_1C_1} \otimes
\vert \Phi^+_i\rangle_{A_2B_2C_2}$ ($i=0,1,2,3$).  $Y_c$ is just the
maximal value of the efficiency in all existing conventional MEPPs
for three-qubit systems \cite{Murao,shengepjd,shengpla}.

As each cross-combination  $\vert\Phi_i^+\rangle \otimes
\vert\Phi_j^+\rangle$ ($i\neq j \in\{0,1,2,3\}$) will lead the three
parties to obtain an entangled two-photon pair, the probability that
the three parties obtain two-photon pairs from a pair of
three-photon systems $P_{3\rightarrow  2}$ is
\begin{eqnarray}
P_{3\rightarrow  2} &=& \sum_{j\neq l=0}^3 F_jF_l\nonumber\\
&=& F_0(F_1 + F_2 + F_3) + F_1(F_0 + F_2 + F_3)\nonumber\\
&+& F_2(F_0 + F_1 + F_3) + F_3(F_0 + F_1 + F_2)\nonumber\\
&=& \frac{2 + 2F_0 -4F_0^2}{3}.
\end{eqnarray}
That is, the efficiency that the three parties obtain three-photon
entangled systems from two-photon entangled systems with
entanglement link $Y_{2\rightarrow  3}$  is
\begin{eqnarray}
Y_{2\rightarrow  3}=\frac{1}{2}P_{3\rightarrow  2}= \frac{1 + F_0 -
2F_0^2}{3}
\end{eqnarray}
because they can obtain a three-photon system from two two-photon
systems.

\begin{figure}[!h]
\begin{center}
\includegraphics[width=8cm,angle=0]{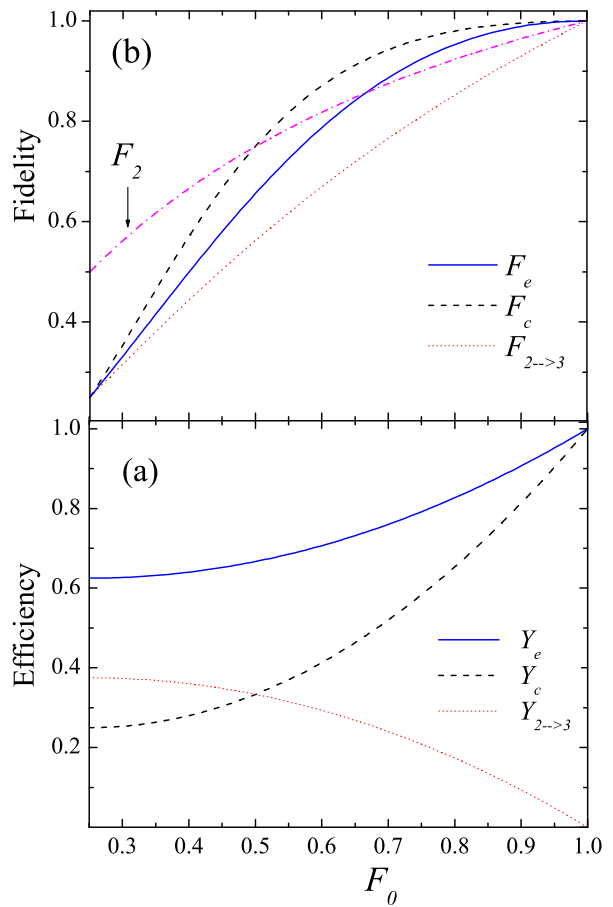}
\caption{(Color online) (a) The efficiency of the present MEPP $Y_e$
and that of the conventional MEPP $Y_c$ (it is just the maximal
value of efficiency from conventional MEPPs) for three-photon
systems under a symmetric noise ($F_1=F_2=F_3$) are shown with a
blue solid line and a black dash line, respectively. Here
$Y_{2\rightarrow 3}$ is the yield that the three parties can obtain
three-photon systems from two-photon systems with entanglement link.
(b) The fidelity of the present MEPP $F_e$ and that of the
conventional MEPP $F_c$. Here $F_2$ and $F_{2\rightarrow 3}$ is the
fidelities of the two-photon systems obtained from the
cross-combinations and that of the three-photon systems obtained
from two-photon systems with entanglement link, respectively. $F_0$
is just the original fidelity of three-photon systems before
entanglement purification.} \label{fig4_efficiency}
\end{center}
\end{figure}

Taking three-photon entanglement production with entanglement link
into account, the efficiency of the present MEPP $Y_e$ is
\begin{eqnarray}
Y_e &=& Y_c + Y_{ 2 \rightarrow 3} = \frac{2- F_0 + 2F^2_0}{3}.
\end{eqnarray}
The efficiency of the present MEPP and the maximal value of that
from conventional MEPPs for three-qubit systems are shown in
Fig.\ref{fig4_efficiency}(a). It is clear that the present MEPP is
more efficient than the conventional MEPPs, especially in the case
that the original fidelity $F_0$ is not big.

The fidelity of our conventional MEPP is
\begin{eqnarray}
F_c &=& \frac{F_0^2}{F^2_0+F^2_1+F^2_2+F^2_3}
= \frac{3F_0^2}{1 - 2F_0 + 4F^2_0}.
\end{eqnarray}
The fidelity of the two-photon systems obtained from
cross-combinations is
\begin{eqnarray}
F_{2} &=& F_0^b=\frac{F_0}{F_0+F_1 }=\frac{3F_0}{1 + 2F_0}.
\end{eqnarray}
The fidelity of the three-photon systems obtained from two-photon
systems with entanglement link is
\begin{eqnarray}
F_{2\rightarrow 3} &=& F_0^t=\frac{F^2_0}{(F_0+F_1)^2
}=\frac{9F^2_0}{1 + 4F_0 + 4F^2_0}.
\end{eqnarray}
The average fidelity of the present MEPP $F_e$ can be calculated as
\begin{eqnarray}
F_{e} &=& (F_cY_c+F_{2\rightarrow 3}Y_{2\rightarrow
3})/Y_e\nonumber\\
&=& \frac{3F^2_0(4 + 7F_0 - 2F^2_0)}{(1 + 2F_0)^2(2 - F_0 +
2F^2_0)}.
\end{eqnarray}
The relation of the four fidelities $F_c$, $F_2$, $F_{2\rightarrow
3}$, and $F_e$ is shown in Fig.\ref{fig4_efficiency}(b).


From Fig.\ref{fig4_efficiency}, one can see that the yield
$Y_{2\rightarrow 3}$ is larger than $Y_c$ when $F_0<\frac{1}{2}$.
$Y_e$ is far larger than $Y_c$, which means that the present MEPP
has a larger efficiency than that in conventional MEPPs. On the
other hand, the fidelity $F_{2\rightarrow 3}$ is smaller than $F_c$
although they both are larger than the original fidelity $F_0$ when
$F_0>\frac{1}{4}$. $F_2$ is larger than $F_c$ when $F_0<\frac{1}{2}$
and it is smaller than $F_c$ when  $F_0>\frac{1}{2}$.

In a practical purification, the three parties need not mix the
three-photon systems obtained by our conventional MEPP and our
recycling MEPP. They can purify them independently in the next
round. Also, they can first purify two-photon systems with the
fidelity $F_2$ and then produce high-fidelity three-photon systems
with entanglement link.

\section{Conventional three-photon entanglement purification for
phase-flip errors}

\label{seciii}

In the process for purifying bit-flip errors, the relative
probabilities of the states $|\Phi_{1}^{\pm}\rangle$,
$|\Phi_{2}^{\pm}\rangle$, and $|\Phi_{3}^{\pm}\rangle$ are
decreased. However, the relative probability of the state
$|\Phi_{0}^{-}\rangle$ is not changed, compared with that of the
state  $|\Phi_{0}^{+}\rangle$. The task of the entanglement
purification for phase-flip errors is in principle to complete the
process with which  the parties can depress the relative probability
of the state $|\Phi_{0}^{-}\rangle$. Same as the entanglement
purification in two-photon systems, a phase-flip error cannot be
corrected directly in three-photon systems, different from a
bit-flip error, but it can be transformed into a bit-flip error with
a Hadamard operation on each photon. That is, after a Hadamard
operation on each photon, the states  $|\Phi_{0}^{+}\rangle$ and
$|\Phi_{0}^{-}\rangle$ shown in Eq.(\ref{GHZstate}) are transformed
into the states $|\Psi^{+}\rangle$ and $|\Psi^{-}\rangle$,
respectively. Here
\begin{eqnarray}
|\Psi^{+}\rangle    =\frac{1}{2}(|HHH\rangle + |HVV\rangle + |VHV\rangle + |VVH\rangle),\nonumber\\
|\Psi^{-}\rangle    =\frac{1}{2}(|HHV\rangle + |HVH\rangle +
|VHH\rangle + |VVV\rangle).\label{phaseflipstate}
\end{eqnarray}

From Eq.(\ref{phaseflipstate}), one can see that the transformation
between phase-flip errors and bit-flip errors in three-photon GHZ
states is more complex than that in Bell states
\cite{Bennett1996,Deutsch,Pan1,Simon,shengpra2008,shengpratwostep,shengpraonestep,lixhepp}.
The three parties cannot use the equipment shown in
Fig.\ref{fig2_three_qubit} to purify the states in
Eq.(\ref{phaseflipstate}) directly. That is, we cannot exploit
simply Hadamard operations to complete the transformation between
phase-flip errors and bit-flip errors in three-photon GHZ states
perfectly, different from Bell states. Fortunately, the number of
the polarization state $\vert V\rangle$ is different in these two
three-photon states. That is,  the number of $|V\rangle$ is even in
the state $|\Psi^{+}\rangle$, while it is odd in the state
$|\Psi^{-}\rangle$. With this feature, the relative probability of
the state $|\Psi^{-}\rangle$ will be depressed with QNDs.

Now, let us use a pair of partner states $|\Psi^{+}\rangle$ and
$|\Psi^{-}\rangle$ as an example to describe the principle of the
purification for phase-flip errors. The density matrix of an
ensemble over noisy channels with only phase-flip errors can be
written as
\begin{eqnarray}
\rho' &=&  p_0|\Psi^{+}\rangle\langle\Psi^{+}| +
p_1|\Psi^{-}\rangle\langle\Psi^{-}|.\label{ensemblerho2}
\end{eqnarray}
Here $p_0$ and $p_1$ represent the probabilities of the states
$|\Psi^{+}\rangle$ and $|\Psi^{-}\rangle$ in the ensemble $\rho'$,
respectively, and $p_0 + p_1=1$. For each pair of the entangled
quantum systems picked out from this ensemble, say $A'_1B'_1C'_1$
and $A'_2B'_2C'_2$, their state can be viewed as the mixture of four
pure states, that is,  $|\Psi^{+}\rangle \otimes |\Psi^{+}\rangle$,
$|\Psi^{+}\rangle \otimes |\Psi^{-}\rangle$, $|\Psi^{-}\rangle
\otimes |\Psi^{+}\rangle$, and $|\Psi^{-}\rangle \otimes
|\Psi^{-}\rangle$ with the probabilities $p_{0}^2$, $p_0p_1$,
$p_1p_0$, and $p_{1}^2$, respectively.

The relation of the outcomes of the parity-check measurements by
Alice, Bob, and Charlie, the states of the quantum system composed
of the six photons $A'_1B'_1C'_1A'_2B'_2C'_2$, and their
probabilities is shown in Table \ref{tab2}. Here
\begin{eqnarray}
|\varphi\rangle_0 &=& \frac{1}{2}(|HHHHHH\rangle+|HVVHVV\rangle \nonumber\\
&& + \;\; |VHVVHV\rangle+|VVHVVH\rangle), 
\end{eqnarray}
\begin{eqnarray}
|\varphi'\rangle_0 &=& \frac{1}{2}(|HHVHHV\rangle+|HVHHVH\rangle  \nonumber\\
&&  + \;\; |VHHVHH\rangle+|VVVVVV\rangle), 
\end{eqnarray}
\begin{eqnarray}
|\varphi\rangle_1 &=& \frac{1}{2}(|HHHHVV\rangle+|HVVHHH\rangle \nonumber\\
&& + \;\; |VHVVVH\rangle+|VVHVHV\rangle),
\end{eqnarray}
\begin{eqnarray}
|\varphi'\rangle_1 &=& \frac{1}{2}(|HHVHVH\rangle+|HVHHHV\rangle  \nonumber\\
&&  + \;\; |VHHVVV\rangle+|VVVVHH\rangle),
\end{eqnarray}
\begin{eqnarray}
|\varphi\rangle_2 &=& \frac{1}{2}(|HHHVHV\rangle+|HVVVVH\rangle \nonumber\\
&& + \;\; |VHVHHH\rangle+|VVHHVV\rangle),
\end{eqnarray}
\begin{eqnarray}
|\varphi'\rangle_2 &=& \frac{1}{2}(|HHVVHH\rangle +  |HVHVVV\rangle \nonumber\\
&&  + \;\; |VHHHHV\rangle + |VVVHVH\rangle), 
\end{eqnarray}
\begin{eqnarray}
|\varphi\rangle_3 &=& \frac{1}{2}(|HHHVVH\rangle + |HVVVHV\rangle \nonumber\\
&& + \;\;|VHVHVV\rangle+|VVHHHH\rangle), 
\end{eqnarray}
\begin{eqnarray}
|\varphi'\rangle_3 &=& \frac{1}{2}(|HHVVVV\rangle +  |HVHVHH\rangle\nonumber\\
&&  + \;\;|VHHHVH\rangle+|VVVHHV\rangle).
\label{evenparity2}
\end{eqnarray}
$U$ in Table \ref{tab2} represents the unitary operations with which
Alice, Bob, and Charlie can transform the six-photon states
$|\varphi\rangle_i$ and $|\varphi'\rangle_i$ ($i=1,2,3$) into the
states $|\varphi\rangle_0$ and $|\varphi'\rangle_0$, respectively.
$I$ and $\sigma_x$ represent the identity operation and the bit-flip
operation, respectively.

One can see that the three parties will obtain the states
$|\varphi\rangle_i$ and $|\varphi'\rangle_i$ ($i=1,2,3$) with the
probabilities $\frac{1}{4}p_0^{2}$ and $\frac{1}{4}p_1^{2}$,
respectively, if the number of odd parities is even. Moreover, the
states $|\varphi\rangle_i$ and $|\varphi'\rangle_i$
 can be transformed into the states $|\varphi\rangle_0$ and
$|\varphi'\rangle_0$ with two local unitary operations,
respectively. That is, if the number of  odd parities is even,
Alice, Bob and Charlie can obtain a six-photon ensemble in which the
probabilities of the states $|\varphi\rangle_0$ and
$|\varphi'\rangle_0$ are $\frac{p_0^{2}}{p_0^{2} + p_1^{2}}$ and
$\frac{p_1^{2}}{p_0^{2} + p_1^{2}}$, respectively.

\begin{table}[!h]
\tabcolsep 0pt \caption{The relation between the outcomes of
parity-check measurements and the six-photon states. } 
\begin{center}
\begin{tabular}{cccc}\hline
  outcomes          \;\;                 & \;\; states                     &   \;\;\;\; probabilities         & \;\;\;\; $U$    \\\hline
  \multirow{2}{*}{even, even, even  }    & \;\; $\vert \varphi\rangle_0$   &   \;\;\;\;  $\frac{1}{4}p_0^2$   &  \multirow{2}{*}{ \;\; $I^{A'_1}\otimes I^{B'_2}\otimes I^{C'_2}$}              \\
                                         & \;\;\; $\vert \varphi'\rangle_0$  &   \;\;\;\;  $\frac{1}{4}p_1^2$               \\\hline
  \multirow{2}{*}{even, odd,  odd   }    & \;\; $\vert \varphi\rangle_1$   &   \;\;\;\;  $\frac{1}{4}p_0^2$   &  \multirow{2}{*}{ \;\; $I^{A'_1}\otimes \sigma_x^{B'_2}\otimes \sigma_x^{C'_2}$  }                  \\
                                         & \;\;\; $\vert \varphi'\rangle_1$  &   \;\;\;\;  $\frac{1}{4}p_1^2$               \\\hline
  \multirow{2}{*}{odd,  even, odd  }     & \;\; $\vert \varphi\rangle_2$   &   \;\;\;\;  $\frac{1}{4}p_0^2$   &  \multirow{2}{*}{ \;\; $\sigma_x^{A'_1}\otimes I^{B'_2}\otimes \sigma_x^{C'_2}$  }                  \\
                                         & \;\;\; $\vert \varphi'\rangle_2$  &   \;\;\;\;  $\frac{1}{4}p_1^2$               \\\hline
  \multirow{2}{*}{odd,  odd,  even  }    & \;\; $\vert \varphi\rangle_3$   &   \;\;\;\;  $\frac{1}{4}p_0^2$   &  \multirow{2}{*}{ \;\; $\sigma_x^{A'_1}\otimes \sigma_x^{B'_2}\otimes I^{C'_2}$ }              \\
                                         & \;\;\; $\vert \varphi'\rangle_3$  &   \;\;\;\;  $\frac{1}{4}p_1^2$
\\\hline
\end{tabular}\label{tab2}
       \end{center}
       \end{table}

Certainly,  Alice, Bob, and Charlie can obtain a three-photon
ensemble in the states $\vert \Psi^+\rangle$ and   $\vert
\Psi^-\rangle$ with the probabilities $\frac{p_0^{2}}{p_0^{2} +
p_1^{2}}$ and $\frac{p_1^{2}}{p_0^{2} + p_1^{2}}$, respectively.
They can obtain this ensemble by performing or not a local
single-photon phase-flip operation on the three-photon systems after
they measure the three photons $A'_2B'_2C'_2$ with the basis
$X=\{\vert \pm\rangle \}$ in each six-photon system
$A'_1B'_1C'_1A'_2B'_2C'_2$. For example, if they obtain the outcomes
of the single-photon measurements $\vert ++-\rangle_{A'_2B'_2C'_2}$
or  $\vert --+\rangle_{A'_2B'_2C'_2}$, they need only perform a
phase-flip operation $\sigma_z=|H\rangle\langle H| -
|V\rangle\langle V|$ on the photon $C'_1$. If they obtain the
outcomes $\vert +-+\rangle_{A'_2B'_2C'_2}$ or  $\vert
-+-\rangle_{A'_2B'_2C'_2}$, they need only perform a phase-flip
operation $\sigma_z=|H\rangle\langle H| - |V\rangle\langle V|$ on
the photon $B'_1$. If the outcomes are $\vert
+--\rangle_{A'_2B'_2C'_2}$ or $\vert -++\rangle_{A'_2B'_2C'_2}$,
they need only perform a phase-flip operation
$\sigma_z=|H\rangle\langle H| - |V\rangle\langle V|$ on the photon
$A'_1$.

With a Hadamard operation on each photon in a three-photon system
kept, the states $\vert \Psi^+\rangle$ and   $\vert \Psi^-\rangle$
become the states $\vert \Phi^+_0\rangle$ and $\vert
\Phi^-_0\rangle$, respectively. That is, by performing the
conventional three-photon entanglement purification for  phase-flip
errors, the three parties can obtain a new ensemble in the states
$\vert \Phi^+_0\rangle$ and $\vert \Phi^-_0\rangle$ with the
probabilities $\frac{p_0^{2}}{p_0^{2}+p_1^{2}}$ and
$\frac{p_1^{2}}{p_0^{2}+p_1^{2}}$, respectively. The fidelity of the
state $\vert \Phi^+_0\rangle$ is
$p'_0=\frac{p_0^{2}}{p_0^{2}+p_1^{2}}=\frac{p_0^{2}}{p_0^{2}+(1-p_0)^{2}}$.
$p'_0>p_0$ when $p_0>1/2$.

The cross-combinations $|\Psi^{+}\rangle\otimes |\Psi^{-}\rangle$
and $|\Psi^{-}\rangle \otimes |\Psi^{+}\rangle$  will lead the
number of odd parities in the outcomes of their parity-check
measurements to be odd and they are discarded as these instances
will lead the three parties to obtain the states $\vert
\Psi^+\rangle$ and $\vert \Psi^-\rangle$ with the same probability
$p_0p_1$. That is, these instances cannot improve the fidelity of
the entangled state $\vert \Psi^+\rangle$ as a phase-flip error in a
GHZ state is an unlocal (global) error, which is different from the
case with some bit-flip errors (a bit-flip error in a GHZ state can
be considered as a local error). The three parties in quantum
communication can only obtain a two-photon ensemble without
entanglement if they want to distil two-photon subsystems from  the
cross-combinations with a phase-flip error.

\section{discussion and summary}

\label{secv}

The present MEPP works for $N$-photon systems in GHZ states, as
shown in detail in Appendix \ref{seciv}.

\begin{figure}[!h]
\begin{center}
\includegraphics[width=7.2cm,angle=0]{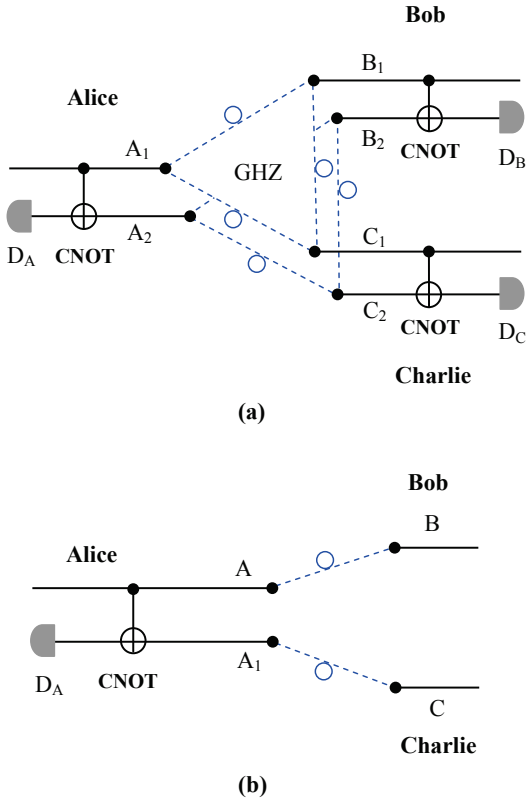}
\caption{(Color online)  The principle of the present multipartite
entanglement purification protocol with CNOT gates. (a) The
conventional entanglement purification for three-photon systems with
CNOT gates; (b) The principle for producing a high-fidelity
three-photon entangled system with entanglement link from two
two-photon subsystems with
 a CNOT gate. }
\label{fig7_three_qubit_cnot}
\end{center}
\end{figure}

It is not difficult to show that the present MEPP works by replacing
the parity-check detectors (QNDs) with CNOT gates.  We can use
three-photon systems as an example to describe the principle of
entanglement purification with entanglement link from two-photon
subsystems, shown in Fig.\ref{fig7_three_qubit_cnot}. The principle
of the conventional three-photon EPP for bit-flip errors and
phase-flip errors with CNOT gates is the same as that in the first
MEPP by Murao \emph{et al.} \cite{Murao}, shown in
Fig.\ref{fig7_three_qubit_cnot}(a). However, our MEPP can also
distil some high-fidelity two-photon subsystems from the
cross-combinations which are just discarded in Ref. \cite{Murao}.
For the cross-combinations  $\vert \Phi^+_0\rangle_{A'B'C'}\otimes
\vert \Phi^+_2\rangle_{A''B''C''}$ and $\vert
\Phi^+_2\rangle_{A'B'C'}\otimes \vert\Phi^+_0\rangle_{A''B''C''}$,
Alice, Bob, and Charlie obtain the three-photon states
$\frac{1}{\sqrt{2}}(\vert HHH\rangle + \vert VVV\rangle)_{A'B'C'}$
and $\frac{1}{\sqrt{2}}(\vert HVH\rangle + \vert
VHV\rangle)_{A'B'C'}$ with the same probability $F_0F_2$ after they
perform their CNOT operations on their two photons and measure the
three photons $A''B''C''$ with the basis $Z$. The outcomes of the
single-photon measurements by Alice, Bob, and Charlie are $\vert
HVH\rangle_{A''B''C''}$ or $\vert VHV\rangle_{A''B''C''}$. After Bob
performs a Hadamard operation on the photon $B'$ and measures it
with the basis $Z$, Alice and Charlie will obtain a two-photon
entangled state $\frac{1}{\sqrt{2}}(\vert HH\rangle + \vert
VV\rangle)_{A'C'}$ with or without a phase-flip operation which
depends on the fact that the outcome of the single-photon
measurement by Bob on the photon $B'$ is $\vert V\rangle_{B'}$ or
$\vert H\rangle_{B'}$, respectively. For the cross-combinations
$\vert \Phi^+_1\rangle_{A'B'C'}\otimes \vert
\Phi^+_3\rangle_{A''B''C''}$ and $\vert
\Phi^+_3\rangle_{A'B'C'}\otimes \vert\Phi^+_1\rangle_{A''B''C''} $,
Alice, Bob, and Charlie also obtain the outcomes $\vert
HVH\rangle_{A''B''C''}$ or $\vert VHV\rangle_{A''B''C''}$. In this
time, they obtain the three-photon states $\frac{1}{\sqrt{2}}(\vert
VHH\rangle + \vert HVV\rangle)_{A'B'C'}$ and
$\frac{1}{\sqrt{2}}(\vert VHH\rangle + \vert HVV\rangle)_{A'B'C'}$
with the same probability $F_1F_3$. After Bob performs a Hadamard
operation on the photon $B'$ and measures it with the basis $Z$,
Alice and Charlie  obtain a two-photon entangled state
$\frac{1}{\sqrt{2}}(\vert HV\rangle + \vert VH\rangle)_{A'C'}$ with
or without a phase-flip operation. That is, when Alice, Bob, and
Charlie obtain the outcomes $\vert HVH\rangle_{A''B''C''}$ or $\vert
VHV\rangle_{A''B''C''}$ after they perform their CNOT operations on
their two photons and measure the three photons $A''B''C''$ with the
basis $Z$, they can obtain a two-photon ensemble in the states
$\frac{1}{\sqrt{2}}(\vert HH\rangle + \vert VV\rangle)_{A'C'}$ and
$\frac{1}{\sqrt{2}}(\vert HV\rangle + \vert VH\rangle)_{A'C'}$ with
the probabilities $F_0F_2$ and $F_1F_3$, respectively. This is just
the case when Alice, Bob, and Charlie obtain their parities (odd,
even, odd) or (even, odd, even) with QNDs discussed in
Sec.\ref{seciic}. So do the other cases. The principle of
entanglement link with CNOT gates is similar to that with QNDs. That
is, the present MEPP works by replacing QNDs with CNOT gates.
However, when the parties replace their QNDs with polarizing beam
splitters, the present MEPP does not work because the
cross-combinations cannot be exploited to distil $N'$-photon
subsystems as there are at least two photons which cannot be
distinguished.

Compared with the conventional MEPPs
\cite{Murao,shengepjd,shengpla}, the present MEPP contains two
parts. One is our conventional MEPP. Its principle is the same as
other conventional MEPPs \cite{Murao,shengepjd,shengpla} although
its efficiency is double as those with QNDs in Refs.
\cite{shengepjd,shengpla} because our conventional MEPP takes all
the instances in which all the parties obtain either an even parity
or an odd parity into account for obtaining high-fidelity
multipartite entangled systems, not only the instances in which all
parties obtain an even parity as those in Refs.
\cite{shengepjd,shengpla}. The other part is our recycling MEPP in
which entanglement link is used to produce a multipartite entangled
system from some subsystems with QNDs. In essence, the parties
distil some multipartite entangled systems from the instances which
are discard in the conventional MEPPs
\cite{Murao,shengepjd,shengpla}, which makes the present MEPP have a
higher efficiency than others.

In the process for describing the principle of our efficient MEPP
with entanglement link from subspaces, we mainly exploit the
cross-Kerr nonlinearity to construct the parity-check detector (that
is,  QND).  We should acknowledge that the implementation of a clean
cross-Kerr nonlinearity is still quite difficult in experiment,
especially  with natural cross-Kerr nonlinearities. It may be
feasible in principle in the present MEPP. On one hand, some works
have been studied on cross-Kerr nonlinearity
\cite{RMP,kok,julio,QND5,discrimination,shapiro1,shapiro2,friedler1,friedler2,friedler3}.
For example, Kok \emph{ et al.} \cite{RMP} showed that operating in
the optical single-photon regime, the Kerr phase shift is only
$\tau\approx10^{-18}$. With electromagnetically induced transparent
materials, cross-Kerr nonlinearities of $\tau\approx10^{-5}$ can be
obtained \cite{kok}. Also, Hofmann \emph{et al.} \cite{QND5} showed
that a phase shift of $\pi$ can be achieved with a single two-level
atom in a one-sided cavity. In 2010, Wittmann \emph{et al.}
\cite{discrimination} investigated quantum measurement strategies
capable of discrimination two coherent states using a homodyne
detector and a photon number resolving detector. On the other hand,
our QND does not require a large nonlinearity and it works for small
values of the cross-Kerr coupling, which makes it possible
\cite{QND1,QND3}.

Although we only exploit the QND based on a cross-Kerr nonlinearity
to explain the principle of our efficient MEPP, other elements can
also be used to construct QNDs, such as quantum dots in optical
cavities, as shown in
Refs.\cite{qndother1,qndother2,qndother3,qndother4,qndother5}. We
use the polarization degree of freedom of photon systems as an
example to describe the principle of our efficient MEPP with
entanglement link from subspaces. Obviously, it works for other
degrees of freedom of quantum systems.

\begin{figure}[!h]
\begin{center}
\includegraphics[width=6cm,angle=0]{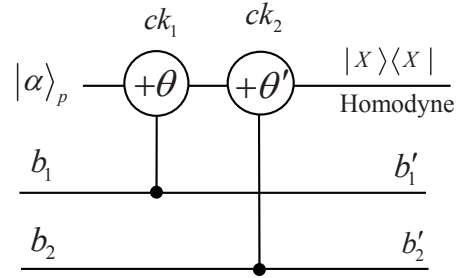}
\caption{The principle of PNDs with cross-Kerr nonlinear media. $+
\theta$  and $+ \theta'$ represent two different cross-Kerr
nonlinear media which introduce the phase shifts $+ \theta$ and $+
\theta'$ when there is a photon passing through the media,
respectively.} \label{fig9_PND}
\end{center}
\end{figure}

Same as all existing EPPs
\cite{Bennett1996,Deutsch,Pan1,Simon,shengpra2008,shengpratwostep,shengpraonestep,lixhepp,dengonestep,Murao,shengepjd,shengpla},
the present MEPP does not take the loss of photons in transmission
over an optical fiber into account. In the principle of the present
scheme, we assume that the parties in quantum communication deal
with their entanglement purification under an ideal condition. That
is, there is no photon loss in transmission. In a practical
transmission of photons over an optical fiber, the losses should be
considered, which will decrease the efficiency of EPPs
\cite{Bennett1996,Deutsch,Pan1,Simon,shengpra2008,shengpratwostep,shengpraonestep,lixhepp,dengonestep,Murao,shengepjd,shengpla},
especially those based on CNOT gates or linear optical elements.
With cross-Kerr nonlinear media, the parties can construct
photon-number detectors (PNDs) to distinguish the number of photons
in the two spatial modes, shown in  Fig.\ref{fig9_PND}. With PNDs,
the present MEPP works efficiently with a photon-loss channel as
well because the parties can in principle determine the number of
the photons in each spatial mode before they perform their MEPP. In
detail, the different phase shifts of the coherent beam $\vert
\alpha\rangle_p$ represent the different numbers of photons passing
through the two spatial modes.

To summarize, we have proposed an efficient MEPP for $N$-photon
systems in a GHZ state.  It contains two parts. One is our
conventional MEPP with which the parties can obtain a high-fidelity
$N$-photon ensemble directly. The other is our recycling MEPP in
which entanglement link is used to produce some $N$-photon entangled
systems from subspaces. Our conventional MEPP is similar to the
conventional MEPP with perfect CNOT gates \cite{Murao}, but it
doubles the efficiency of the MEPP with QNDs based on cross-Kerr
nonlinearity in Ref. \cite{shengepjd} and the MEPP for electronic
systems \cite{shengpla}. In the present MEPP, the parties can obtain
not only high-fidelity $N$-photon entangled systems directly but
also high-fidelity $N'$-photon entangled subsystems ($2 \leq N'<N$).
With entanglement link, the parties can produce some high-fidelity
$N$-photon entangled systems from $N'$-photon entangled subsystems.
In the entanglement purification for phase-flip errors, the parties
can only perform the conventional process as a phase-flip error in
GHZ states is an unlocal one. The present MEPP has higher efficiency
than the conventional MEPPs \cite{Murao,shengepjd,shengpla} as the
cross-combinations, which are just the discarded instances in the
latter, can be used to distil high-fidelity $N'$-photon  entangled
subsystems. We discuss the principle of our MEPP with the
entanglement link in detail for the three-photon systems. The result
can be generalized to the case with $N$-photon systems. Moreover,
the present MEPP works by replacing parity-check detectors with CNOT
gates or replacing the polarization degree of freedom  of $N$-photon
systems with others.

\section*{ACKNOWLEDGEMENTS}

This work is supported by the National Natural Science Foundation of
China under Grant Nos. 10974020 and 11174039, A Foundation for the
Author of National Excellent Doctoral Dissertation of P. R. China
under Grant No. 200723,  Beijing Natural Science Foundation under
Grant No. 1082008, and the Fundamental Research Funds for the
Central Universities.

\appendix

\section{High-efficiency multipartite entanglement purification with entanglement link}

\label{seciv}

\subsection{Conventional $N$-photon entanglement purification for bit-flip errors and  phase-flip errors}
\label{seciva}

The GHZ state of a multipartite entangled system composed of $N$
two-level particles can be described as
\begin{eqnarray}
|\Phi^{+}_0\rangle_N=\frac{1}{\sqrt{2}}(|HH\cdots H\rangle +
|VV\cdots V\rangle)_{A,B,\cdots, Z}. \label{state2}
\end{eqnarray}
Here the subscripts $A$, $B$, $\cdots$ and  $Z$ represent the
photons sent to the parties Alice, Bob, $\cdots$, and Zach,
respectively. Certainly, there are another $2^N-1$ GHZ states for an
$N$-qubit system and can be written as
\begin{eqnarray}
|\Phi^{+}_{ij\cdots k}\rangle_N=\frac{1}{\sqrt{2}}(|ij\cdots
k\rangle + |\bar{i}\bar{j}\cdots \bar{k}\rangle)_{AB\cdots C}
\end{eqnarray}
and
\begin{eqnarray}
|\Phi^{-}_{ij\cdots k}\rangle_N=\frac{1}{\sqrt{2}}(|ij\cdots
k\rangle - |\bar{i}\bar{j}\cdots \bar{k}\rangle)_{AB\cdots
C}.\end{eqnarray} Here $\bar{i}=1-i$, $\bar{j}=1-j$,  $\bar{k}=1-k$,
and $i,j,k\in \{0,1\}$. $\vert 0\rangle\equiv \vert H\rangle$ and
$\vert 1\rangle\equiv \vert V\rangle$.

For correcting the bit-flip errors in multipartite entangled quantum
systems, we can also divide the whole entanglement purification into
two steps. One is the conventional multipartite entanglement
purification and the other is the entanglement purification with
entanglement link from subspaces. The conventional entanglement
purification for multipartite entangled quantum systems with
bit-flip errors is similar to that for three-photon entangled
quantum systems. We should only increase the number of  the QNDs and
the Hadamard operations shown in Fig. \ref{fig2_three_qubit}. Let us
use a simple example to describe the principle of the conventional
entanglement purification for $N$-photon systems. That is, let us
assume that the ensemble of photon systems after the transmission
over a noisy channel is in the state
\begin{eqnarray}
\rho_N&=&f_0|\Phi^+_0\rangle_N\langle \Phi^+_0| +\cdots +
f_{ij\cdots k} |\Phi^+_{ij\cdots k}\rangle_N\langle \Phi^+_{ij\cdots
k}| \nonumber\\ &&  + \cdots +
f_{2^{N-1}-1}|\Phi^+_{2^{N-1}-1}\rangle_N\langle
\Phi^+_{2^{N-1}-1}|. \label{stateNm}
\end{eqnarray}
Here $f_{ij\cdots k}$ presents the probability that an $N$-photon
system is in the state $|\Phi^+_{ij\cdots k}\rangle_N$ and
\begin{eqnarray}
f_0 + \cdots + f_{ij\cdots k} + \cdots + f_{2^{N-1}-1}=1.
\end{eqnarray}
In conventional multipartite entanglement purification for each two
$N$-photon systems, the parties in quantum communication will keep
the identity-combinations
 $|\Phi^+_0\rangle_N \otimes |\Phi^+_0\rangle_N$, $\cdots$,
$|\Phi^+_{ij\cdots k}\rangle_N \otimes |\Phi^+_{ij\cdots
k}\rangle_N$, $\cdots$, and $|\Phi^+_{2^{N-1}-1}\rangle_N \otimes
|\Phi^+_{2^{N-1}-1}\rangle_N $ with the probabilities $f^2_0$,
$\cdots$, $f_{ij\cdots k}^2$, $\cdots$, and $f_{2^{N-1}-1}^2$,
respectively. That is, they keep the instances in which they all
obtain the even parity and those in which they all obtain the odd
parity with their QNDs.

When all the parties obtain the even parity, the system with $2N$
photons is in the state $\vert \phi_{ij\cdots k}\rangle_{2N}=
\frac{1}{\sqrt{2}}(|ij\cdots k\rangle|ij\cdots k\rangle +
|\bar{i}\bar{j}\cdots \bar{k}\rangle|\bar{i}\bar{j}\cdots
\bar{k}\rangle)_{A_1,B_1,\cdots, Z_1,A_2,B_2,\cdots, Z_2}$ with the
probability  $\frac{1}{2}f_{ij\cdots k}^2$. When they all obtain the
odd parity, the system  is in the state $\vert \psi_{ij\cdots
k}\rangle_{2N} = \frac{1}{\sqrt{2}}(|ij\cdots
k\rangle|\bar{i}\bar{j}\cdots \bar{k}\rangle  +
|\bar{i}\bar{j}\cdots \bar{k}\rangle |ij\cdots
k\rangle)_{A_1,B_1,\cdots, Z_1,A_2,B_2,\cdots, Z_2}$
 with the probability
$\frac{1}{2}f_{ij\cdots k}^2$. Certainly, the parties can transform
the states $\vert \psi_{ij\cdots k}\rangle_{2N}$ ($i,j, k=0,1,$)
into the states $\vert \phi_{ij\cdots k}\rangle_{2N}$ with a
bit-flip operation $\sigma_x=\vert H\rangle\langle V\vert + \vert
V\rangle\langle H\vert$ on each of the $N$ photons $A_2$, $B_2$,
$\cdots$, and $Z_2$.

By measuring the photons $A_2$, $B_2$, $\cdots$, and $Z_2$ with the
basis $X=\{\vert \pm\rangle\}$ and performing some unitary
operations or not, the parties can obtain a new  $N$-photon system
which is in the states $\vert
\Phi_0\rangle_{N}=\frac{1}{\sqrt{2}}(\vert HH\cdots H\rangle   +
\vert VV\cdots V\rangle)_{A_1,B_1,\cdots, Z_1}$, $\cdots$, $\vert
\Phi_{ij\cdots k}\rangle_{N}=\frac{1}{\sqrt{2}}(|ij\cdots k\rangle +
|\bar{i}\bar{j}\cdots \bar{k})_{A_1,B_1,\cdots, Z_1}$, $\cdots$, and
$\vert \Phi_{2^{N-1}-1}\rangle_{N}= \frac{1}{\sqrt{2}}(\vert
HV\cdots V\rangle  + \vert VH\cdots H\rangle)_{A_1,B_1,\cdots,
Z_1,A_2,B_2,\cdots, Z_2}$ with the probabilities $ f^2_0$, $\cdots$,
$ f_{ij\cdots k}^2$, $\cdots$, and $f_{2^{N-1}-1}^2$, respectively,
similar to  the conventional entanglement purification for
three-photon systems shown in Sec.\ref{seciib}. In this way, the
parties can obtain a  new ensemble of $N$-photon systems $\rho'_N$
with the fidelity $f'_0=\frac{f^2_0}{f^2_0 + \cdots + f_{ij\cdots
k}^2 + \cdots + f_{2^{N-1}-1}^2}$ from the original ensemble in the
state $\rho_N$.

In the conventional $N$-photon entanglement purification for
phase-flip errors, the parties can only obtain a new ensemble,
similar to the case for three-photon systems. In detail, with a
phase-flip error on the state $|\Phi^{+}_0\rangle_N$, it becomes
$|\Phi^{-}_0\rangle_N$. Here
\begin{eqnarray}
|\Phi^{-}_0\rangle_N=\frac{1}{\sqrt{2}}(|HH\cdots H\rangle
-|VV\cdots V\rangle)_{A,B,\cdots, Z}. \label{statenphase}
\end{eqnarray}
With a Hadamard operation on each photon, the states
$|\Phi_{0}^{+}\rangle_N$ and $|\Phi_{0}^{-}\rangle_N$  are
transformed into the states $|\Psi_0^{+}\rangle_N$ and
$|\Psi_0^{-}\rangle_N$, respectively. Here
\begin{eqnarray}
|\Psi^{+}_0\rangle_N &=& \frac{1}{2^{\frac{N+1}{2}}}[(|H\rangle +
|V\rangle)_A (\vert  H\rangle + \vert V\rangle)_B \cdots (\vert
H\rangle + \vert V\rangle)_Z \nonumber\\
&+& (\vert H\rangle - \vert V\rangle)_A (\vert H\rangle -\vert
V\rangle)_B \cdots (\vert H\rangle -\vert
V\rangle)_Z],\nonumber\\
|\Psi^{-}_0\rangle_N &=& \frac{1}{2^{\frac{N+1}{2}}}[(|H\rangle +
|V\rangle)_A (\vert  H\rangle + \vert V\rangle)_B \cdots (\vert
H\rangle + \vert V\rangle)_Z \nonumber\\
&-& (\vert H\rangle - \vert V\rangle)_A (\vert H\rangle -\vert
V\rangle)_B \cdots (\vert H\rangle -\vert
V\rangle)_Z].\nonumber
\end{eqnarray}
The number of $|V\rangle$ is even in each item of the state
$|\Psi_0^{+}\rangle_N$, while it is odd in the state
$|\Psi_0^{-}\rangle_N$.  With this feature, the relative probability
of the state $|\Psi_0^{-}\rangle_N$ will be depressed.

The density matrix of an ensemble over noisy channels with only
phase-flip errors can be written as
\begin{eqnarray}
\rho'_N &=&  P_0|\Psi_0^{+}\rangle_N\langle\Psi_0^{+}| +
P_1|\Psi_0^{-}\rangle_N\langle\Psi_0^{-}|.\label{ensemblerho2}
\end{eqnarray}
Here $P_0$ and $P_1$ represent the probabilities of the states
$|\Psi_0^{+}\rangle_N$ and $|\Psi_0^{-}\rangle_N$ in the ensemble
$\rho_N'$, respectively, and $P_0 + P_1=1$. For each pair of the
entangled $N$-photon systems picked out from this ensemble, say
$A_1B_1\cdots Z_1$ and $A_2B_2\cdots Z_2$, their state can be viewed
as the mixture of four pure states, that is,  $|\Psi_0^{+}\rangle_N
\otimes |\Psi_0^{+}\rangle_N$, $|\Psi_0^{+}\rangle_N \otimes
|\Psi_0^{-}\rangle_N$, $|\Psi_0^{-}\rangle_N \otimes
|\Psi_0^{+}\rangle_N$, and $|\Psi_0^{-}\rangle_N \otimes
|\Psi_0^{-}\rangle_N$ with the probabilities $P_{0}^2$, $P_0P_1$,
$P_1P_0$, and $P_{1}^2$, respectively.

The parties let their photons pass through their QNDs, as the same
as those shown in Fig.\ref{fig2_three_qubit}. That is, Alice lets
her photons $A_1$ and $A_2$ pass through her QND. So do the other
parties.  The parties only keep the instances in which the number of
odd parities is even when they perform a parity-check measurement on
their two photons independently. In these instances, the photons
come from the state   $|\Psi_0^{+}\rangle_N \otimes
|\Psi_0^{+}\rangle_N$ or $|\Psi_0^{-}\rangle_N \otimes
|\Psi_0^{-}\rangle_N$. It is not difficult to prove that the
instances coming from the cross-combinations $|\Psi_0^{-}\rangle_N
\otimes |\Psi_0^{+}\rangle_N$ and $|\Psi_0^{+}\rangle_N \otimes
|\Psi_0^{-}\rangle_N$  lead the number of odd parities to be odd. By
performing the conventional $N$-photon entanglement purification for
phase-flip errors, similar to the case for three-photon systems, the
$N$  parties can obtain a new ensemble in the states $\vert
\Phi^+_0\rangle_N$ and $\vert \Phi^-_0\rangle_N$ with the
probabilities $\frac{P_0^{2}}{P_0^{2}+P_1^{2}}$ and
$\frac{P_1^{2}}{P_0^{2}+P_1^{2}}$, respectively. The fidelity of the
state $\vert \Phi^+_0\rangle_N$ is
$P'_0=\frac{P_0^{2}}{P_0^{2}+P_1^{2}}=\frac{P_0^{2}}{P_0^{2}+(1-P_0)^{2}}$.
When $P_0>1/2$, $P'_0>P_0$.

\subsection{High-efficiency multipartite entanglement purification
for  bit-flip errors  with entanglement link} \label{secivb}

In the conventional multipartite entanglement purification for
bit-flip errors, the parties do not take the cross-combinations
$|\Phi^+_{lr\cdots q}\rangle_N \otimes |\Phi^+_{ij\cdots
k}\rangle_N$ and $ |\Phi^+_{ij\cdots k}\rangle_N \otimes
|\Phi^+_{lr\cdots q}\rangle_N$ ($l,r,\cdots, q \in \{0,1\}$ and
$l\neq i$, $r\neq j$, $\cdots$, or $q\neq k$) into account for
obtaining some high-fidelity $N$-photon entangled systems. However,
the parties can first obtain some high-fidelity $N'$-photon
entangled systems ($2\leq N'<N$) from the cross-combinations and
then obtain some high-fidelity $N$-photon entangled systems with
entanglement link, similar to the entanglement purification for
three-photon entangled systems (as discussed in Sec.\ref{seciic} and
Sec. \ref{seciid}). In this time, the probability that the
$2N$-photon system is in the  cross-combination $|\Phi^+_{lr\cdots
q}\rangle_N \otimes |\Phi^+_{ij\cdots k}\rangle_N$ or $
|\Phi^+_{ij\cdots k}\rangle_N \otimes |\Phi^+_{lr\cdots q}\rangle_N$
is $f_{lr\cdots q}f_{ij\cdots k}$. The high-fidelity entangled
subsystem which can be obtained by the $N$ parties is composed of
$N'_s$ photons. Here $N'_s$ is a integer number and it is larger
than $\frac{N-1}{2}$ but not larger than $\frac{N+1}{2}$. Certainly,
the more the number of the photons in each system, the more the
kinds of the entanglement purification with entanglement link. Let
us use four-photon systems
 as an example to describe the principle of
high-efficiency multipartite entanglement purification with
entanglement link. It is more complicated than that in the case with
three-photon systems.

For four-photon systems, the sixteen GHZ states can be written as
\begin{eqnarray}
|\Phi_{0}^{\pm}\rangle_{ABCD}=\frac{1}{\sqrt{2}}(|HHHH\rangle \pm |VVVV\rangle)_{ABCD},\nonumber\\
|\Phi_{1}^{\pm}\rangle_{ABCD}=\frac{1}{\sqrt{2}}(|HHHV\rangle \pm |VVVH\rangle)_{ABCD},\nonumber\\
|\Phi_{2}^{\pm}\rangle_{ABCD}=\frac{1}{\sqrt{2}}(|HHVH\rangle \pm |VVHV\rangle)_{ABCD},\nonumber\\
|\Phi_{3}^{\pm}\rangle_{ABCD}=\frac{1}{\sqrt{2}}(|HHVV\rangle \pm |VVHH\rangle)_{ABCD},\nonumber\\
|\Phi_{4}^{\pm}\rangle_{ABCD}=\frac{1}{\sqrt{2}}(|HVHH\rangle \pm
|VHVV\rangle)_{ABCD},\nonumber\\
|\Phi_{5}^{\pm}\rangle_{ABCD}=\frac{1}{\sqrt{2}}(|HVHV\rangle \pm
|VHVH\rangle)_{ABCD},\nonumber\\
|\Phi_{6}^{\pm}\rangle_{ABCD}=\frac{1}{\sqrt{2}}(|HVVH\rangle \pm
|VHHV\rangle)_{ABCD},\nonumber\\
|\Phi_{7}^{\pm}\rangle_{ABCD}=\frac{1}{\sqrt{2}}(|HVVV\rangle \pm
|VHHH\rangle)_{ABCD}. \label{GHZstate4}
\end{eqnarray}
Here the subscripts $A$, $B$, $C$, and $D$ represent the photons
kept by the parties in quantum communication Alice, Bob, Charlie,
and Dean, respectively.  When we only consider the entanglement
purification for bit-flip errors, we assume that the ensemble of
four-photon systems after the transmission over a noisy channel is
in the state
\begin{eqnarray}
\rho_4&=& \sum_{m=0}^{7}f''_m|\Phi^+_m\rangle_{ABCD}\langle
\Phi^+_m|. \label{stateNm4}
\end{eqnarray}

Except for the 8 identity-combinations
$|\Phi^+_m\rangle_{A_1B_1C_1D_1}\otimes
 |\Phi^+_m\rangle_{A_2B_2C_2D_2}$ kept for obtaining some four-photon systems with a high fidelity, the 56 cross-combinations
$|\Phi^+_{lr\cdots q}\rangle_4 \otimes |\Phi^+_{ij\cdots
k}\rangle_4$ are discarded in the conventional four-photon
entanglement purification. In fact, the parties can first distil
some three-photon entangled systems and two-photon entangled systems
from the cross-combinations, and then obtain some four-photon
entangled systems with entanglement link.

When the parities of the four parties with QNDs are even, even,
even, and odd (we abbreviate them as (even, even, even, odd) below),
the 8-photon system $A_1B_1C_1D_1A_2B_2C_2D_2$ is in the states
\begin{eqnarray}
\vert \zeta_1 \rangle &=& \frac{1}{\sqrt{2}}(\vert
HHHH\rangle_{A_1B_1C_1D_1}\vert
HHHV\rangle_{A_2B_2C_2D_2} \nonumber\\
&& \;\;\;\; +\; \vert VVVV\rangle_{A_1B_1C_1D_1}\vert
VVVH\rangle_{A_2B_2C_2D_2}), \nonumber\\
\vert \zeta_2 \rangle &=& \frac{1}{\sqrt{2}}(\vert
HHHV\rangle_{A_1B_1C_1D_1}\vert HHHH\rangle_{A_2B_2C_2D_2} \nonumber\\
&& \;\;\;\; +\; \vert VVVH\rangle_{A_1B_1C_1D_1}\vert
VVVV\rangle_{A_2B_2C_2D_2}), \nonumber\\
\vert \zeta_3 \rangle  &=& \frac{1}{\sqrt{2}}(\vert
HHVH\rangle_{A_1B_1C_1D_1}\vert HHVV\rangle_{A_2B_2C_2D_2} \nonumber\\
&& \;\;\;\; +\; \vert VVHV\rangle_{A_1B_1C_1D_1}\vert
VVHH\rangle_{A_2B_2C_2D_2}), \nonumber\\
\vert \zeta_4 \rangle &=& \frac{1}{\sqrt{2}}(\vert
HHVV\rangle_{A_1B_1C_1D_1}\vert HHVH\rangle_{A_2B_2C_2D_2} \nonumber\\
&& \;\;\;\; +\; \vert VVHH\rangle_{A_1B_1C_1D_1}\vert
VVHV\rangle_{A_2B_2C_2D_2}), \nonumber\\
\vert \zeta_5 \rangle &=& \frac{1}{\sqrt{2}}(\vert
HVHH\rangle_{A_1B_1C_1D_1}\vert HVHV\rangle_{A_2B_2C_2D_2} \nonumber\\
&& \;\;\;\; +\; \vert VHVV\rangle_{A_1B_1C_1D_1}\vert
VHVH\rangle_{A_2B_2C_2D_2}), \nonumber\\
\vert \zeta_6 \rangle  &=& \frac{1}{\sqrt{2}}(\vert
HVHV\rangle_{A_1B_1C_1D_1}\vert HVHH\rangle_{A_2B_2C_2D_2} \nonumber\\
&& \;\;\;\; +\; \vert VHVH\rangle_{A_1B_1C_1D_1}\vert
VHVV\rangle_{A_2B_2C_2D_2}), \nonumber\\
\vert \zeta_7 \rangle &=& \frac{1}{\sqrt{2}}(\vert
HVVH\rangle_{A_1B_1C_1D_1}\vert HVVV\rangle_{A_2B_2C_2D_2} \nonumber\\
&& \;\;\;\; +\;
\vert VHHV\rangle_{A_1B_1C_1D_1}\vert VHHH\rangle_{A_2B_2C_2D_2}), \nonumber
\end{eqnarray}
and
\begin{eqnarray}
\vert \zeta_8 \rangle &=& \frac{1}{\sqrt{2}}(\vert
HVVV\rangle_{A_1B_1C_1D_1}\vert HVVH\rangle_{A_2B_2C_2D_2} \nonumber\\
&& \;\;\;\; +\;
\vert VHHH\rangle_{A_1B_1C_1D_1}\vert VHHV\rangle_{A_2B_2C_2D_2}) \nonumber
\end{eqnarray}
with the probabilities of
 $\frac{1}{2}f''_0f''_1$, $\frac{1}{2}f''_0f''_1$, $\frac{1}{2}f''_2f''_3$, $\frac{1}{2}f''_2f''_3$, $\frac{1}{2}f''_4f''_5$,
  $\frac{1}{2}f''_4f''_5$, $\frac{1}{2}f''_6f''_7$, and
$\frac{1}{2}f''_6f''_7$, respectively.

The four parties in quantum communication can obtain a three-photon
system by performing a Hadamard operation on each of the five
photons $D_1A_2B_2C_2D_2$ and then measuring them with the basis
$Z$. When the number of the outcomes $\vert V\rangle$  in the
measurements is even, the three-photon system $A_1B_1C_1$ is in the
states $\vert \Phi^+_0\rangle_{A_1B_1C_1}$,  $\vert
\Phi^+_1\rangle_{A_1B_1C_1}$, $\vert \Phi^+_2\rangle_{A_1B_1C_1}$,
and  $\vert \Phi^+_3\rangle_{A_1B_1C_1}$ with the probabilities
$\frac{1}{2}f''_0f''_1$, $\frac{1}{2}f''_2f''_3$,
$\frac{1}{2}f''_4f''_5$, and  $\frac{1}{2}f''_6f''_7$, respectively.
When the number of the outcomes $\vert V\rangle$  is odd, the system
$A_1B_1C_1$ is in the states $\vert \Phi^-_0\rangle_{A_1B_1C_1}$,
$\vert \Phi^-_1\rangle_{A_1B_1C_1}$, $\vert
\Phi^-_2\rangle_{A_1B_1C_1}$, and  $\vert
\Phi^-_3\rangle_{A_1B_1C_1}$ with the probabilities
$\frac{1}{2}f''_0f''_1$, $\frac{1}{2}f''_2f''_3$,
$\frac{1}{2}f''_4f''_5$, and  $\frac{1}{2}f''_6f''_7$, respectively.
With a phase-flip operation $\sigma_z$ on one of the three photons
$A_1B_1C_1$, the parties can obtain the state $\vert
\Phi^+_m\rangle_{A_1B_1C_1}$ from the state $\vert
\Phi^-_m\rangle_{A_1B_1C_1}$ ($m=0,1,2,3$). That is, the three
parties Alice, Bob, and Charlie obtain a three-photon ensemble in
the state
\begin{eqnarray}
\rho''_{ABC}&=& \sum_{m=0}^{3}f'''_m|\Phi^+_m\rangle_{ABC}\langle
\Phi^+_m|, \label{stateNm43}
\end{eqnarray}
where
\begin{eqnarray}
f'''_0 &=& \frac{f''_0f''_1}{f''_0f''_1 + f''_2f''_3 + f''_4f''_5 +
f''_6f''_7},\nonumber\\
f'''_1 &=& \frac{f''_2f''_3}{f''_0f''_1 + f''_2f''_3 + f''_4f''_5 +
f''_6f''_7},\nonumber\\
f'''_2 &=& \frac{f''_4f''_5}{f''_0f''_1 + f''_2f''_3 + f''_4f''_5 +
f''_6f''_7},\nonumber\\
f'''_3 &=& \frac{f''_6f''_7}{f''_0f''_1 + f''_2f''_3 + f''_4f''_5 +
f''_6f''_7}.
\end{eqnarray}

When the parities are (even, even, odd, even), the three parties
Alice, Bob, and Dean obtain a three-photon ensemble $ \rho''_{ABD}$,
similar to the case with the outcomes (even, even, even, odd). So do
the ensembles $ \rho''_{ACD}$ and $ \rho''_{BCD}$.

When two of the four parties obtain the odd parity with their QNDs,
they can obtain a two-photon ensemble $\rho''_{AB}$, $\rho''_{AC}$,
$\rho''_{AD}$, $\rho''_{BC}$, $\rho''_{BD}$ or $\rho''_{CD}$, by
performing some Hadamard operations and measurements with the basis
$Z$. Let us use the outcomes (even, even, odd, odd) as an example to
describe the principle. In this time, the 8-photon system is in the
states
\begin{eqnarray}
\vert \xi_1 \rangle &=& \frac{1}{\sqrt{2}}(\vert
HHHH\rangle_{A_1B_1C_1D_1}\vert
HHVV\rangle_{A_2B_2C_2D_2} \nonumber\\
&& \;\;\;\; +\; \vert VVVV\rangle_{A_1B_1C_1D_1}\vert
VVHH\rangle_{A_2B_2C_2D_2}), \nonumber\\
\vert \xi_2 \rangle &=& \frac{1}{\sqrt{2}}(\vert
HHVV\rangle_{A_1B_1C_1D_1}\vert HHHH\rangle_{A_2B_2C_2D_2} \nonumber\\
&& \;\;\;\; +\; \vert VVHH\rangle_{A_1B_1C_1D_1}\vert
VVVV\rangle_{A_2B_2C_2D_2}), \nonumber\\
\vert \xi_3 \rangle &=& \frac{1}{\sqrt{2}}(\vert
HHHV\rangle_{A_1B_1C_1D_1}\vert HHVH
\rangle_{A_2B_2C_2D_2} \nonumber\\
&& \;\;\;\; +\; \vert VVVH\rangle_{A_1B_1C_1D_1}\vert
VVHV\rangle_{A_2B_2C_2D_2}), \nonumber\\
\vert \xi_4 \rangle &=& \frac{1}{\sqrt{2}}(\vert
HHVH\rangle_{A_1B_1C_1D_1}\vert HHHV
\rangle_{A_2B_2C_2D_2} \nonumber\\
&& \;\;\;\; +\; \vert VVHV\rangle_{A_1B_1C_1D_1}\vert
VVVH\rangle_{A_2B_2C_2D_2}), \nonumber\\
\vert \xi_5 \rangle &=& \frac{1}{\sqrt{2}}(\vert
HVHH\rangle_{A_1B_1C_1D_1}\vert HVVV
\rangle_{A_2B_2C_2D_2} \nonumber\\
&& \;\;\;\; +\; \vert VHVV\rangle_{A_1B_1C_1D_1}\vert
VHHH\rangle_{A_2B_2C_2D_2}), \nonumber\\
\vert \xi_6 \rangle &=& \frac{1}{\sqrt{2}}(\vert
HVVV\rangle_{A_1B_1C_1D_1}\vert HVHH
\rangle_{A_2B_2C_2D_2} \nonumber\\
&& \;\;\;\; +\; \vert VHHH\rangle_{A_1B_1C_1D_1}\vert
VHVV\rangle_{A_2B_2C_2D_2}), \nonumber\\
\vert \xi_7 \rangle &=& \frac{1}{\sqrt{2}}(\vert
HVHV\rangle_{A_1B_1C_1D_1}\vert HVVH
\rangle_{A_2B_2C_2D_2} \nonumber\\
&& \;\;\;\; +\; \vert VHVH\rangle_{A_1B_1C_1D_1}\vert
VHHV\rangle_{A_2B_2C_2D_2}), \nonumber
\end{eqnarray}
and
\begin{eqnarray}
\vert \xi_8 \rangle &=& \frac{1}{\sqrt{2}}(\vert
HVVH\rangle_{A_1B_1C_1D_1}\vert HVHV
\rangle_{A_2B_2C_2D_2} \nonumber\\
&& \;\;\;\; +\; \vert VHHV\rangle_{A_1B_1C_1D_1}\vert
VHVH\rangle_{A_2B_2C_2D_2}) \nonumber
\end{eqnarray}
with the probabilities of
 $\frac{1}{2}f''_0f''_3$,
 $\frac{1}{2}f''_0f''_3$,
 $\frac{1}{2}f''_1f''_2$,
 $\frac{1}{2}f''_1f''_2$,
 $\frac{1}{2}f''_4f''_7$,
 $\frac{1}{2}f''_4f''_7$, $\frac{1}{2}f''_5f''_6$, and $\frac{1}{2}f''_5f''_6$, respectively. For
obtaining two-photon entangled systems, Alice, Bob, Charlie, and
Dean first perform a Hadamard operation on each of the six photons
$C_1D_1A_2B_2C_2D_2$ and then measure them with the basis $Z$. When
the number of the outcomes $\vert V\rangle$ is even, the two-photon
system $A_1B_1$ is in the states $\vert
\phi^+\rangle_{A_1B_1}=\frac{1}{\sqrt{2}}(\vert HH\rangle + \vert
VV\rangle)_{A_1B_1}$ and  $\vert
\psi^+\rangle_{A_1B_1}=\frac{1}{\sqrt{2}}(\vert HV\rangle + \vert
VH\rangle)_{A_1B_1}$  with the probabilities $\frac{1}{2}(f''_0f''_3
+ f''_1f''_2)$ and $\frac{1}{2}(f''_4f''_7 + f''_5f''_6)$,
respectively.  When the number of the outcomes $\vert V\rangle$ is
odd, the two-photon system $A_1B_1$ is in the states $\vert
\phi^-\rangle_{A_1B_1}=\frac{1}{\sqrt{2}}(\vert HH\rangle - \vert
VV\rangle)_{A_1B_1}$ and  $\vert
\psi^-\rangle_{A_1B_1}=\frac{1}{\sqrt{2}}(\vert HV\rangle - \vert
VH\rangle)_{A_1B_1}$  with the probabilities $\frac{1}{2}(f''_0f''_3
+ f''_1f''_2)$ and $\frac{1}{2}(f''_4f''_7 + f''_5f''_6)$,
respectively. With a phase-flip operation $\sigma_z$ on the photon
$A_1$, the states $\vert \phi^-\rangle_{A_1B_1}$ and  $\vert
\psi^-\rangle_{A_1B_1}$ will be transformed into the states $\vert
\phi^+\rangle_{A_1B_1}$ and  $\vert \psi^+\rangle_{A_1B_1}$,
respectively. That is, Alice and Bob obtain a two-photon ensemble
 in the state
\begin{eqnarray}
\rho''_{AB}&=& f^0_{AB}|\phi^+\rangle_{AB}\langle \phi^+| +
f^1_{AB}|\psi^+\rangle_{AB}\langle \psi^+|, \label{stateNm43}
\end{eqnarray}
where
\begin{eqnarray}
f^0_{AB} &=& \frac{f''_0f''_3 + f''_1f''_2}{f''_0f''_3 +
f''_1f''_2 + f''_4f''_7 + f''_5f''_6},\nonumber\\
f^1_{AB} &=& \frac{f''_4f''_7 + f''_5f''_6}{f''_0f''_3 + f''_1f''_2
+ f''_4f''_7 + f''_5f''_6}.
\end{eqnarray}

\begin{figure}[!h]
\begin{center}
\includegraphics[width=6cm,angle=0]{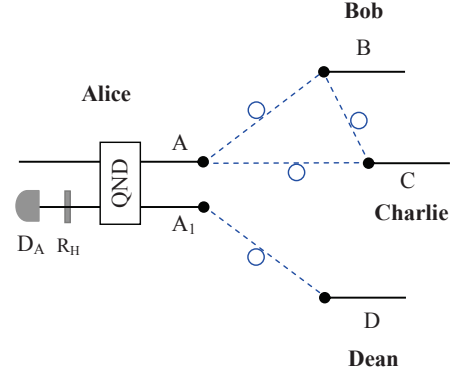}
\caption{(Color online)  The principle of the entanglement link for
producing a four-photon entangled system from a three-photon
entangled subsystem and a two-photon entangled subsystem with a QND.
} \label{fig_two_three_qubit}
\end{center}
\end{figure}

\begin{figure}[!h]
\begin{center}
\includegraphics[width=7.2cm,angle=0]{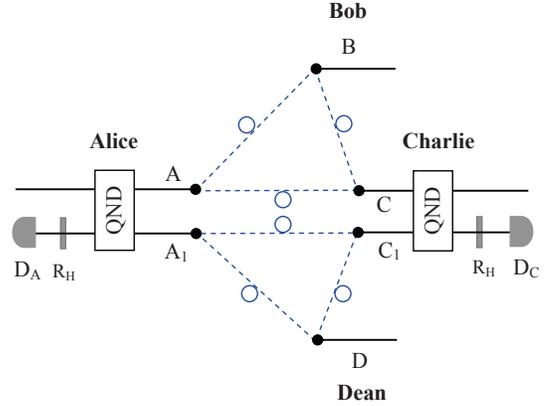}
\caption{(Color online)  The principle of the entanglement link for
producing a four-photon entangled system from two three-photon
entangled subsystems  with two QNDs. } \label{fig_three_three_qubit}
\end{center}
\end{figure}

\begin{figure}[!h]
\begin{center}
\includegraphics[width=7.2cm,angle=0]{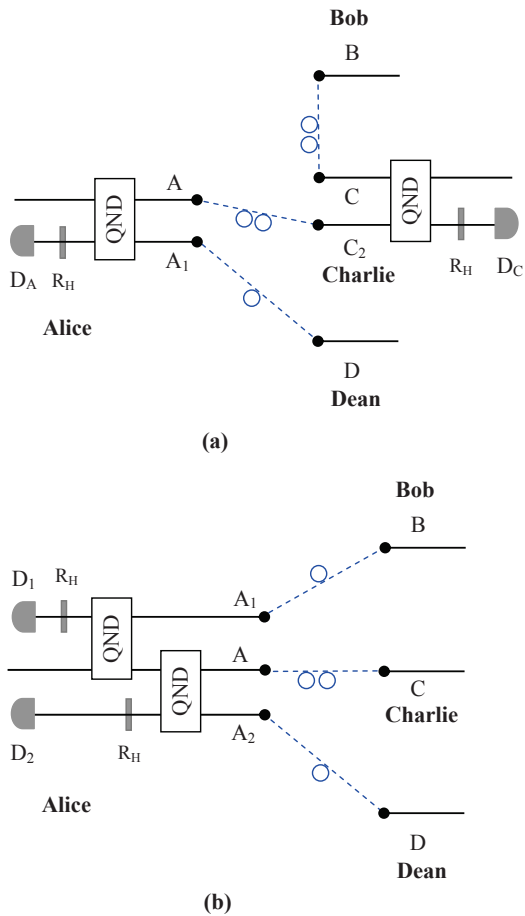}
\caption{(Color online)  The principle of the entanglement link for
producing a four-photon entangled subsystem from three two-photon
entangled subsystems  with two QNDs. There are two topological
structures: (a) a symmetrical structure in which Alice shares a
two-photon entangled system with Charlie and another with Dean, and
Charlie shares another two-photon entangled system with Bob;  (b) an
unsymmetrical structure in which Alice shares a two-photon entangled
system with each of the other three parties. }
\label{fig_two_two_qubit1}
\end{center}
\end{figure}

With three-photon ensembles $ \rho''_{ABC}$, $ \rho''_{ABD}$,  $
\rho''_{ACD}$, and $ \rho''_{BCD}$, and two-photon ensembles
$\rho''_{AB}$, $\rho''_{AC}$, $\rho''_{AD}$, $\rho''_{BC}$,
$\rho''_{BD}$, and $\rho''_{CD}$,  Alice, Bob, Charlie, and Dean can
obtain some four-photon systems with entanglement link. In detail,
for a system composed of a three-photon entangled subsystem $ABC$
and a two-photon entangled subsystem $A_1D$, Alice, Bob, Charlie,
and Dean can obtain a four-photon entangled system $ABCD$ by
performing a QND measurement on the photons $AA_1$ and performing a
single-photon measurement on the photon $A_1$ with the basis $Z$
after Alice takes a Hadamard operation on the photon $A_1$,  as
shown in Fig. \ref{fig_two_three_qubit}. Certainly, they can obtain
a four-photon entangled system $ABCD$ from the complicated system
composed of two three-photon subsystems $ABC$ and $A_1C_2D$ by
performing a QND measurement on the photons $A$  and $A_1$, and
another on the photons $C$ and $C_1$, shown in Fig.
\ref{fig_three_three_qubit}. By performing a single-photon
measurement on the photon $A_1$ and another on the photon $C_1$
after a Hadamard operation on each of these two photons, Alice, Bob,
Charlie, and Dean can obtain a four-photon entangled system. Also,
the four parties can obtain a four-photon entangled systems from
three two-photon entangled subsystems, shown in Fig.
\ref{fig_two_two_qubit1}.

\end{document}